\documentclass[twocolumn,epjc3]{svjour3}          

\RequirePackage[T1]{fontenc}
\smartqed  

\RequirePackage{graphicx}
\RequirePackage{epstopdf}
\RequirePackage{bbding}
\RequirePackage{lmodern}
\RequirePackage[colorlinks,citecolor=blue,urlcolor=blue,linkcolor=blue]{hyperref}
\RequirePackage{amssymb}
\RequirePackage{slashed}
\RequirePackage{amsmath}
\RequirePackage{subfigure}
\RequirePackage{bbding}
\hypersetup{pdfstartview=}
\journalname{Eur. Phys. J. C}

\makeatother

\begin{document}

\title{Neutrino charge radius and electromagnetic dipole moments via scalar and vector leptoquarks}
\author{A. Bola\~nos--Carrera    \thanksref{e1,addr1}\and
        M. Guiot--Lomel\'i  \thanksref{e2,addr2}\and
        G. Tavares-Velasco  \thanksref{e3,addr2}
}

\thankstext{e1}{susy\_musik@hotmail.com}
\thankstext{e2}{marianne.gl93@gmail.com}
\thankstext{e3}{gtv@fcfm.buap.mx}

\institute{Tecnologico de Monterrey, School of Engineering and Sciences, Avenida Eugenio Garza Sada 2501, CP 64849, Monterrey, Nuevo Leon, Mexico.\label{addr1}\and 
Facultad de Ciencias F\'isico-Matem\'aticas,\\
  Benem\'erita Universidad Aut\'onoma de Puebla,\\
 C.P. 72570, Puebla, Pue., Mexico \label{addr2} }

\date{Received: date / Accepted: date}
\maketitle

\begin{abstract}
The one-loop contribution of  scalar and vector leptoquarks (LQs) to the electromagnetic properties (NEPs) of massive Dirac neutrinos  is presented via  an effective Lagrangian approach. For the contribution of gauge LQs to the effective neutrino charge radius defined in \cite{Bernabeu:2002nw}, we considered a Yang-Mills scenario and used the background field method for the calculation.  Analytical results for  nonzero neutrino mass are presented, which can be useful to obtain the NEPs of heavy neutrinos, out of which approximate expressions are obtained for light neutrinos. For the numerical analysis we concentrate on the only renormalizable scalar and vector LQ representations that do not need  extra symmetries to forbid tree-level proton decay. Constraints on the parameter space consistent with current experimental data are then discussed and  it is found that the scalar  $\widetilde{R}_2$ and the vector  $U_1$ representations could yield the largest LQ contributions to the NEPs: for LQ  couplings to both left- and right handed neutrinos of the order of $O(1)$ and a LQ mass of $1.5$-$2$ TeV, the magnetic dipole moment (MDM) of a Dirac neutrino with a mass in the eV scale can be of the order of $10^{-9}$-$10^{-10}$ $\mu_B$, whereas its neutrino electric dipole moment (EDM) can reach values as high as $10^{-20}$-$10^{-19}$ ecm.  On the other hand,  the effective NCR can reach values up to $10^{-35}$ cm$^2$ even if LQs do not couple to right-handed neutrinos, in which case  the EDM vanishes and the contribution to the MDM  would be negligible. 
\end{abstract}

\section{Introduction}

The neutrino interaction with the photon  is governed by the neutrino electromagnetic properties (NEPs) \cite{Bernstein:1963jp,Fujikawa:1980yx,Shrock:1982sc,Kayser:1982br}, which can only arise at the one-loop level or higher orders in perturbation theory and have been long the focus of considerable interest in the literature \cite{Giunti:2008ve,Giunti:2014ixa,Wong:2005pa,Broggini:2012df} since they could allow us to determine the Dirac or Majorana nature of neutrinos \cite{Kayser:1982br} and also hint new physics effects in neutrino experiments\cite{Broggini:2012df,Studenikin:2019bmw,Studenikin:2020nky,Giunti:2015gga}.
The neutrino electromagnetic vertex function consistent with Lorentz and electromagnetic gauge invariance can be written as \cite{Fujikawa:2003ww}
\begin{align}
\Gamma^\mu_{A\bar{\nu}\nu}&=ie\Big(\gamma^\mu\left(F_1^V(q^2)+F_1^A(q^2)\gamma^5\right)\nonumber\\&+\frac{i}{2m_\nu}\sigma^{\mu\nu}q_\nu\left(F^V_2(q^2)+F^A_2(q^2)\gamma^5\right)\nonumber\\&+q^\mu\left(F^V_3(q^2)+F^A_3(q^2)\gamma^5\right)\Big),
\end{align}
where $q^2$ is the photon squared transfer momentum,  $F_1^V$ is the electric charge form factor and $F_1^A$  the anapole form factor. As for the chirality-flipping  form factors $F_2^{V}$  and $F_2^{A}$, they determine the static $CP$-conserving magnetic dipole moment (MDM) $\mu_\nu$ and the static $CP$-violating electric dipole moment (EDM) $d_\nu$ as follows
\begin{equation}
\mu_{\nu}=e\frac{F_2^V(0)}{2m_\nu},
\end{equation}
and
\begin{equation}
d_{\nu}=-ie\frac{F_2^A(0)}{2m_\nu}.
\end{equation}
Both dipole form factors vanish for a Majorana neutrino, which can only have nonvanishing anapole form factor and transition dipole form factors \cite{Kayser:1982br,Shrock:1982sc,Schechter:1981hw}. On the other hand, Dirac neutrino can have nonvanishing anapole and dipole form factors \cite{Mohapatra:1974gc,Shrock:1982sc}. Thus, the observation of a neutrino MDM would be a clear evidence that neutrinos are Dirac particles. 

The standard model (SM) augmented with massive Dirac neutrinos predicts a non-vanishing neutrino MDM  at the one-loop level, which, in units of the Bohr magneton, $\mu_B=e/(2m_e)$ is
given by \cite{Fujikawa:1980yx,Shrock:1982sc}
\begin{equation}
\mu_{\nu_\ell}=\frac{3 eG_F m_{\nu_\ell}}{8\sqrt{2}\pi^2}\simeq 3.1\times 10^{-19}\left(\frac{m_{\nu_\ell}}{1\,{\rm eV}}\right)\, \mu_B,
\end{equation}
whereas the EDM   vanishes at the one-loop level, though transition EDMs can be nonvanishing.

Although the neutrino electric charge form factor $F_1^V(q^2)$  vanishes for an on shell photon, except in theories with electrically millicharged neutrinos \cite{Giunti:2014ixa}, it can give rise to a nonvanishing neutrino charge radius (NCR): 
\begin{equation}
\langle r^2 \rangle_{\nu_\ell}=6\frac{dF_1^V(q^2)}{dq^2}\Big|_{q^2=0}.
\end{equation}
Using the conventional definition of $F_1^V(q^2)$, this quantity was calculated long ago in the context of the SM \cite{Fujikawa:1972fe,Bardeen:1972vi,Lee:1977tib,Monyonko:1984gb,Lucio:1984jn} and was found to be gauge dependent, which  stems from the fact that off shell Green's functions are not associated with $S$-matrix elements and can thus be plagued with several pathologies \cite{Binosi:2009qm}, such as dependence on the gauge-fixing parameter (GFP)  $\xi$,  ultraviolet divergences , etc.  
Nonetheless,  well behaved off shell Green's functions, out of which physical observables may be defined, can be obtained via  the pinch technique (PT) \cite{Cornwall:1981zr,Cornwall:1989gv,Papavassiliou:1989zd,Binosi:2009qm}. This approach  was followed by the authors of Refs. \cite{Papavassiliou:1989zd,Bernabeu:2000hf,Bernabeu:2002nw,Bernabeu:2002pd}, who addressed all the theoretical and experimental issues  to define an effective NCR that is finite, GFP independent, and target independent, thereby being a valid physical observable that can serve as a probe of the SM at neutrino experiments, as discussed in \cite{Bernabeu:2002pd}. 

The PT is a diagrammatic approach that requires to insert the associated off shell vertex  into a physical process and judiciously removing the gauge-dependent terms arising from box, vertex, and self-energy diagrams. 
Furthermore, it has been shown that, at least  up to the two-loop level, the well-behaved off shell  Green's functions obtained through the PT are identical to those obtained via the background field method (BFM), as long as  the Feynman-'t Hooft gauge is used \cite{Denner:1994xt,Pilaftsis:1996fh}. This provides a systematic approach to obtain new physics contributions to the effective NCR. 

\begin{figure*}[!hbt]
\centering\includegraphics[width=14cm]{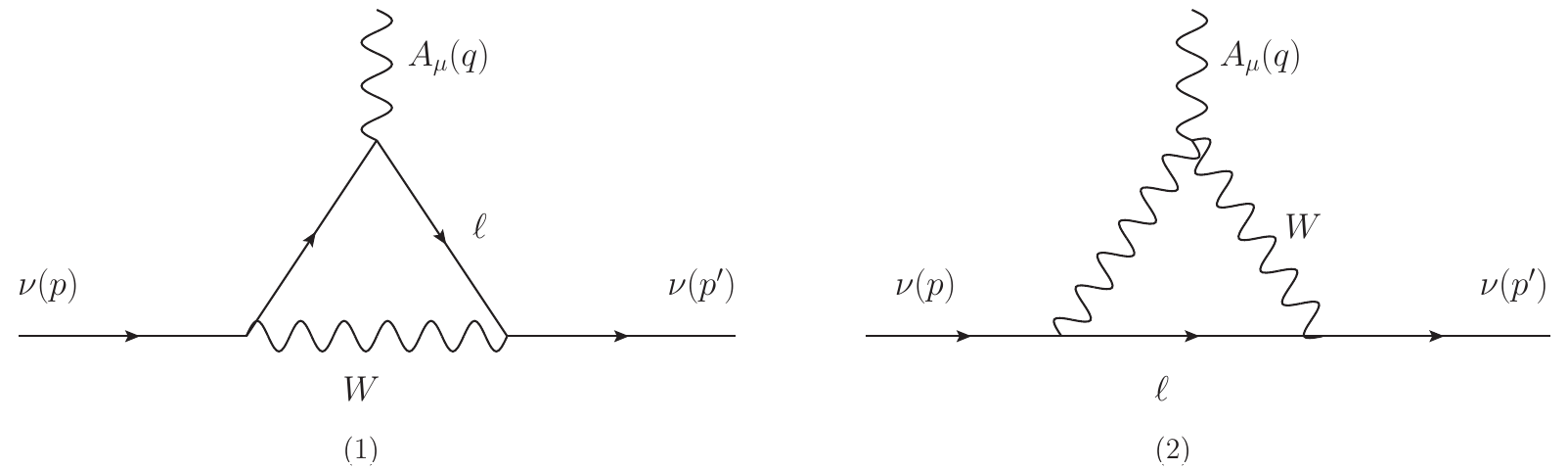}
\caption{Feynman diagrams that contribute to leading order in the charged lepton mass to the effective NCR at the one-loop level in the SM in the BFM. \label{SMNCR}}
\end{figure*}

The correspondence between the effective NCR obtained via the PT and that obtained through the BFM was  discussed in Ref. \cite{Bernabeu:2002pd} in the context of the SM, where the one-loop level  contributions to the effective NCR are induced by the Feynman diagrams of Fig. \ref{SMNCR} to leading order  in the mass of the charged lepton $\ell$.

The corresponding result is \cite{Bernabeu:2000hf,Bernabeu:2002pd,Papavassiliou:2003rx}

\begin{align}
\label{SMNCRRes}
\langle r^2 \rangle_{\nu_\ell}&=\frac{G_F}{4\sqrt{2}\pi^2}\left(3-2\log\left(\frac{m_{\ell}^2}{m_W^2}\right)\right)
\nonumber\\&\simeq\left\{
\begin{array}{lcr}
4.1\times 10^{-33}\;\text{cm}^2&\text{for}\,&\nu_e,\\
2.4\times 10^{-33}\;\text{cm}^2&\text{for}\,&\nu_\mu,\\
1.5\times 10^{-33}\;\text{cm}^2&\text{for}\,&\nu_\tau.\\
\end{array}\right.
\end{align}

It is worth mentioning that the $Z-\gamma$ self-energy does not yield a contribution to the effective NCR as it is associated to the effective electroweak charge form factor rather than the electric charge one, as discussed in  \cite{Bernabeu:2000hf}.  

In this work we present a calculation of the one-loop contributions to neutrino MDM, EDM and effective NCR in scalar and vector leptoquark (LQ) models.
LQs are hypothetical spin-0 or spin-1 particles  predicted by several new physics theories \cite{Pati:1974yy,Georgi:1974sy,Georgi:1974yf,Fritzsch:1974nn,Dimopoulos:1980hn,Senjanovic:1982ex,Frampton:1989fu,Schrempp:1984nj,Buchmuller:1985nn,Gripaios:2009dq,Witten:1985xc,Hewett:1988xc,Ellis:1980hz,Farhi:1980xs,Hill:2002ap}, which are  peculiar as  they carry both lepton and baryon number, thereby interacting simultaneously to leptons and quarks and giving rise to a rich phenomenology. Such particles were first predicted in the $SU(4)_R\times SU(4)_L\times SU(4')$ model of Pati and Salam \cite{Pati:1974yy}, which postulates that lepton number is the fourth color quantum number, but they also appear naturally in grand unification theories (GUT) \cite{Georgi:1974sy,Georgi:1974yf,Fritzsch:1974nn,Dimopoulos:1980hn,Senjanovic:1982ex,Frampton:1989fu},    theories with composite fermions \cite{Schrempp:1984nj,Buchmuller:1985nn,Gripaios:2009dq}, superstring-inspired $E_6$ models \cite{Witten:1985xc,Hewett:1988xc}, technicolor models \cite{Ellis:1980hz,Farhi:1980xs,Hill:2002ap}, etc. LQs can give rise to new interesting effects, out of which the most dramatic is the appearance of tree-level LQ-diquark couplings that can trigger rapid proton decay \cite{Langacker:1980js} and constrain severely the LQ mass unless ad-hoc symmetries are invoked to preserve proton stability.

The late 1990s saw a boom of interest in LQs promp\--ted by the apparent anomaly in $e^+p$ scattering reported  by the H1 \cite{H1:1997utt}  and ZEUS \cite{ZEUS:1997bxs} collaborations, which could be explained by the presence of LQ particles, but such an interest faded away once no further confirmation of a SM deviation was found in subsequent analyses \cite{ZEUS:1999pjj}. Very recently, however,  the interest in LQs has renewed  as they can explain  the apparent  lepton flavor universality  violating (LFUV) anomalies in semi-leptonic $b$-hadron decays, namely the $R_{D,D^*}$ and $R_{K,K^*}$ anomalies,  but can also provide a solution to the muon $g-2$ discrepancy: a plethora of LQ models constructed to this end have been proposed \cite{Marzocca:2021azj,Angelescu:2021lln,Bauer:2015knc,Becirevic:2016yqi,Kumar:2018kmr,Cheung:2022zsb,Bhaskar:2022vgk,Saad:2020ihm,Saad:2020ucl,DaRold:2020bib,BhupalDev:2020zcy,Altmannshofer:2020axr,Fuentes-Martin:2020bnh,Gherardi:2020qhc,Bigaran:2020jil,Dorsner:2020aaz,Babu:2020hun,Crivellin:2020tsz}.
Even more, LQs can generate small neutrino masses radiatively  \cite{AristizabalSierra:2007nf,Saad:2020ihm,Babu:2020hun,Zhang:2021dgl,Cai:2017wry,Popov:2016fzr,Babu:2020hun}. Although the latest LHC-b data  seem to exclude the $R_{K,K^*}$ anomaly \cite{LHCb:2022qnv}, it is still worth studying the LQ effects on experimental observables. The  neutrino MDM has already been calculated in the framework of LQ models \cite{Hewett:1997ba,Chua:1998yk,Povarov:2007zz,Sanchez-Velez:2022nwm}, but an up-to-date analysis is in order given the recent proposal of  LQ models. As far as LQ contributions to both the EMD and the effective NCR,  no known calculation exists yet to our knowledge.


In the experimental arena,  bounds on the neutrino MDM and NCR already exist. We list in Table \ref{NEPbounds} the  90\% CL limits obtained from several experiments. Other recent bounds such as those obtained \cite{Khan:2022akj} by using the COHERENT experiment data  \cite{COHERENT:2021xmm} are not shown as they are  less stringent.

\begin{table*}[!htb]
\caption{90\% CL  upper bounds (or allowed range) on neutrino MDM and NCR for each flavor from several neutrino experiments: XENONnT \cite{Khan:2022bel}, Borexino \cite{Borexino:2017fbd}, GEMMA \cite{Beda:2012zz}, DONUT \cite{DONUT:2001zvi}, LSND \cite{LSND:2001akn}, TEXONO \cite{TEXONO:2009knm},  and solar \cite{Khan:2017djo}.}
\label{NEPbounds}
\begin{center}
\begin{tabular}{ccc}
\hline \hline
Flavor & $|\mu _{\nu }|\ (\times 10^{-11}\mu _{B})$ & $\left \langle r_{\nu}^{2}\right \rangle \ \big(\times 10^{-32}$cm$^{2}\big)$
\\ \hline\hline
$\nu _{e}$ & \multicolumn{1}{c}{$%
\begin{array}{l}
 0.63\ \text{(XENONnT)}  \ \\
 3.9\  \text{(Borexino)}\  \\
  2.9\  \text{(GEMMA)}
\end{array}%
$} & \multicolumn{1}{c}{$
\begin{array}{l}
[-45,\ 3.0] \ \text{(XENONnT)} \\
\lbrack -0.82,\ 1.27]\  \text{(Solar)} \\
\lbrack -5.94,\ 8.28]\  \text{(LSND)} \\
\lbrack -4.2,\ 6.6]\  \  \text{(TEXONO)}
\end{array}%
$}\\  \hline
$\nu _{\mu}$ &\multicolumn{1}{c}{$%
\begin{array}{l}
 1.37\ \text{(XENONnT)}\ \\
 5.8\  \text{(Borexino)}\ \\
\end{array}%
$}& \multicolumn{1}{l}{$%
\begin{array}{l}
[-45,\ 52] \ \text{(XENONnT)} \  \\
\lbrack -9,\ 31]\  \text{(Solar)} \\
 1.2\  \text{(CHARM-II)} \\
\lbrack -4.2,\ 0.48]\  \text{(TEXONO)}%
\end{array}%
$}\\ \hline
$\nu _{\tau }$ & \multicolumn{1}{c}{$%
\begin{array}{l}
 1.24\times 10^{4}\ \text{(XENONnT)} \ \\
 5.8\times 10^{4}\  \text{(Borexino)} \\
  3.9\times 10^{4}\ \text{(DONUT)}%
\end{array}%
$} &\multicolumn{1}{c}{$
\begin{array}{l}
[-40,\ 45] \ \text{(XENONnT)} \\
\lbrack -9,\ 31]\  \text{(Solar)}
\end{array}%
$}  \\ \hline \hline
\end{tabular}%
\\[0pt]
\end{center}
\end{table*}
As far as the neutrino EDM, no experimental limit exist yet, but several indirect and theoretical bounds have been obtained. For the electron and muon neutrino EDMs, the most stringent limits  are $d_{\nu_e,\nu_\mu}\lesssim 10^{-21}$ ecm \cite{delAguila:1990jg}, whereas  for the tau neutrino  bounds  of the order of $10^{-17}-10^{-18}$ ecm have been obtained in the context of some new physics scenarios: $d_{\nu_\tau}\lesssim \times 10^{-17}$ ecm in a model-independent approach (naturalness) \cite{Akama:2001fp}, $d_{\nu_\tau}\lesssim \times 10^{-17}$ ecm in effective Lagrangians  \cite{Escribano:1996wp}, $d_{\nu_\tau}\lesssim 10^{-18}-10^{-20}$ ecm in \ vector-like multiplet models   \cite{Ibrahim:2010va}, etc. Also, it was found \cite{Llamas-Bugarin:2017upv} that the ILC and CLIC would allow one to test $d_{\nu_\tau}$ values up to the $10^{-19}$ ecm level for center-of-mass energies of $500$ to $3000$ GeV.

The rest of this work is organized as follows. In Section \ref{Model} we present an overview of some minimal renormalizable models that predict   scalar and vector LQs at the TeV scale and have no proton decay at the tree-level. Section \ref{Calculation} is devoted to present the calculation of the electromagnetic properties of a massive Dirac neutrino arising from both vector and scalar LQs, whereas Sec. \ref{NumAnal} is devoted to the discussion of the current constraints on the  LQ parameter space from experimental data along with a numerical estimate of the NEPs of SM neutrinos. Finally, the conclusion and outlook are presented in Sec. \ref{Summary}. The lengthy formulas for the analytical results of the LQ contributions to the NEPs are presented in  \ref{AnalyRes} and some results for the LQ contributions to several observables useful to constrain the LQ parameter space are presented in  \ref{lowenerobser}.

\section{Minimal models with  scalar and vector LQs}
\label{Model}
We are interested in the electromagnetic properties of massive Dirac neutrinos and assume that there are right-handed neutrinos that could interact with LQ particles and SM quarks but are sterile to the weak force. We will not focus on the mechanism of neutrino mass generation, though there are several LQ models that address this problem \cite{AristizabalSierra:2007nf,Saad:2020ihm,Babu:2020hun,Zhang:2021dgl,Cai:2017wry,Popov:2016fzr,Babu:2020hun}.
Instead of analyzing a particular LQ model in all its complexity, it is more convenient to study the LQ effects on low-energy experiments in a model-independent fashion via an effective Lagrangian.
Apart from $SU(3)_C\times SU(2)_L\times U(1)_Y$ gauge invariance, extra symmetries must be assumed to forbid dangerous diquark couplings that can induce rapid proton decay at tree-level, thereby pushing the LQ masses up to the ultra high-energy scale: for instance, proton decay sets a limit of $10^{16}$ GeV on the mass of GUT vector  LQs \cite{Langacker:1980js,Georgi:1974yf}.
Along these lines, the authors  of Ref. \cite{Buchmuller:1986zs} found all the  $SU(3)_C\times SU(2)_L\times U(1)_Y$ LQ representations with renormalizable couplings to SM fermion bilinear operators respecting both baryon and lepton number conservation: it turns out that there are only five  scalar and five vector LQ representations of such a kind \cite{Buchmuller:1986zs}, though there are one extra  scalar  and one extra vector representations when right-handed neutrinos are included \cite{Davies:1990sc,Dorsner:2016wpm}. Out of all these twelve LQ representations, only ten of them can couple to neutrinos, thereby yielding one-loop contributions to NEPs. Such representations  are shown in Table \ref{LQrep}, where apart from the spin, fermion number, and $SU(3)_c\times SU(2)_L\times U(1)_Y$ gauge quantum numbers of each LQ representation, we also include the charge of the corresponding LQs and indicate which of them can couple to left- and/or right-handed neutrinos.

\begin{table*}[!htb]
\caption{$SU(3)_c\times SU(2)_L\times U(1)_Y$ LQ representations with renormalizable couplings to  bilinear fermion operators, including right-handed neutrinos,  respecting both baryon and lepton number conservation \cite{Buchmuller:1986zs,Davies:1990sc,Dorsner:2016wpm}.  We show  the corresponding spin $s$,  fermion number $F=3B+L$, and   $SU(3)_c\times SU(2)_L\times U(1)_Y$ gauge quantum numbers (GQNs). We also include the electric charge of the LQ components of each representation and indicate between parenthesis if they  couple to left-handed $(L)$ neutrinos, right-handed $(R)$ neutrinos, or both of them (LR)
\cite{Dorsner:2016wpm}. In the last column we show the representations that provide renormalizable LQ models that do not need to invoke extra symmetries to forbid proton decay in perturbation theory.  \label{LQrep}}
\begin{center}
\begin{tabular}{cccccc}
\hline
\hline
Representation&$s$&$F$& GQNs&LQ electric charge ($|e|$)&Extra symmetries required to forbid proton decay\\
\hline
\hline
$S_3$&$0$&$-2$&$(\bar{3},3,1/3)$&$1/3$ $(L)$, $4/3$, $-2/3$ $(L)$&Yes\\
$R_2$&$0$&$0$&$(3,2,7/6)$&$5/3$, $2/3$ $(L)$&No\\
$\widetilde{R}_2$&$0$&$0$&$(3,2,1/6)$&$2/3$ $(R)$, $-1/3$ $(LR)$&No\\
$S_1$ &$0$&$-2$&$(\bar{3},1,1/3)$&$1/3$ $(LR)$&Yes\\
$\bar{S}_1$&$0$&$-2$&$(\bar{3},1,-2/3)$&$-2/3$ $(R)$&Yes\\
$U_3$&$1$&$0$&$(3,3,2/3)$&$2/3$  $(L)$, $-1/3$  $(L)$, $5/3$&No\\
$V_2$&$1$&$-2$&$(\bar{3},2,5/6)$&$1/3$  $(L)$, $4/3$&Yes\\
$\widetilde{V}_2$ &$1$&$-2$&$(\bar{3},2,-1/6)$&$1/3$  $(R)$, $-2/3$  $(LR)$&Yes\\
$U_1$&$1$&$0$&$(3,1,2/3)$&$2/3$  $(LR)$&No\\
$\bar{U}_1$&$1$&$0$&$(3,1,-1/3)$&$-1/3$  $(R)$&Yes\\
\hline
\hline
\end{tabular}
\end{center}
\end{table*}

Another approach was followed by the authors of Refs. \cite{Arnold:2013cva,Assad:2017iib}, who found the effective LQ models based on  $SU(3)_c\times SU(2)_L\times U(1)_Y$  representations with renormalizable couplings to fermion bilinear operators that do not need to invoke any extra global symmetry to forbid baryon number violation in perturbation theory, thereby forbidding  tree-level proton decay  via either diquark couplings or triple and quadruple LQ self-interactions. These models are thus still phenomenologically viable at the TeV scale as they have no severe constraints on the LQ masses and couplings from proton decay. It was found that the only models of this kind are those comprised by  the either one or several of the following four LQ representations: the two scalar representations $R_2$ and $\widetilde{R}_2$  \cite{Arnold:2013cva} and the two vector representations $U_1$ and $U_3$ \cite{Assad:2017iib}. Models with one or several of these four LQ representations have been the focus of attention recently  since, apart from providing a renormalizable framework and predicting a rich phenomenology at the TeV scale, they can explain the LFUV anomalies in $B$-meson decays as well as the muon $g-2$ anomaly  in regions of the parameter space still allowed by the  current experimental constraints from  meson decays, electroweak precision parameters, and direct searches at the LHC \cite{Fajfer:2015ycq,Calibbi:2017qbu,Crivellin:2018yvo,Datta:2019bzu,Hati:2019ufv,Hati:2019ufv,DaRold:2019fiw,Cornella:2019hct,Angelescu:2021lln,Bhaskar:2021pml,Cheung:2022zsb}. 

Although the $R_2$, $\widetilde{R}_2$, $U_1$, and $U_3$ representations can  give contributions to NEPs,  only $\widetilde{R}_2$ and $U_1$  can have couplings to both left- and right-handed neutrinos.  
In particular, the $U_1$ representation has been the source of attention recently as  emerges naturally from  the minimal realization of the Pati-Salam model \cite{Pati:1974yy} and provides a solution to the LFUV anomalies in $B$-meson decays \cite{Bordone:2017bld,Bordone:2018nbg,Barbieri:2015yvd,Buttazzo:2017ixm,Angelescu:2018tyl,Cornella:2019hct,Becirevic:2016oho,Bhattacharya:2016mcc,DiLuzio:2017vat,Aebischer:2022oqe,Assad:2017iib,Heeck:2018ntp,Calibbi:2017qbu,Aydemir:2018cbb}, for which models based on the  $\widetilde{R}_2$ have also been studied \cite{Chen:2022hle,Dey:2017ede,Becirevic:2016oho,Becirevic:2016yqi}.
Notice however that the scalar ${S}_1$ and  vector $\widetilde{V}_2$  representations, with $F=2$, can also can have couplings to both left- and right-handed neutrinos, so 
in order to have a comprehensive  calculation we will consider the scalar representations $\widetilde{R}_2$ and $S_1$ as well as the vector representations  $U_1$ and  $\widetilde{V}_2$ since they provide the most general LQ couplings to neutrinos and quarks that can induce contributions to NEPs at the one-loop level. 

We will present below an overview of the LQ couplings necessary for our calculation.

\subsection{Scalar LQ interactions}
\subsubsection{$\widetilde{R}_2$ and $S_1$ neutrino couplings}
In the following, we will refrain from 
presenting the LQ interactions with quarks and charged leptons as they are not relevant for our calculation and can be found elsewhere \cite{Buchmuller:1986zs,Dorsner:2016wpm}. Therefore, the dimension-4 Yukawa lagrangian for the $\widetilde{R}_2$ and $S_1$ representations that yield LQ interactions  with a quark and both left- and right-handed neutrinos   can be written as \cite{Buchmuller:1986zs,Dorsner:2016wpm}
\begin{align}
\mathcal{L}_{\widetilde{R}_2}&\supset
-\widetilde{Y}^{RL}_{2\,ij}  \overline{d}^{'i}_R i \widetilde{R}_2^T\tau_2 L_L^{'j}+
\widetilde{Y}^{LR}_{2\,ij} \overline{Q}^{'i}_L  \widetilde{R}_2 \nu^{'j}_R+{\rm H.c.},
\end{align}
and
\begin{align}
\mathcal{L}_{S_1}&\supset Y^{LL}_{1\, ij}\overline{Q'}^{C \,i}_L i\tau_2L^{'j}_L S_1+Y^{\overline{RR}}_{1\,ij}\overline{d}^{'C\,i}_R\nu^{'j}_R S_1
+{\rm H.c.},
\end{align}
where as customary $L_L^{'i}$ and $Q_L^{'i}$ are $SU(2)_L$ left-handed lepton and quark doublets, respectively, whereas $\nu_{R}^{'i}$ and $q_{R}^{'i}$ are $SU(2)$ singlets, with  the subscripts $i$ and $j$ being family indices . The left-side (right-side) superscript of the Yukawa matrices $Y$  stand for the chirality of the corresponding quark (lepton) multiplet.

We now write the LQ in terms of their components and  rotate to the mass eigenstates of the fermions via the transformations $e^{i'}_L=e^{i}_L$, $d^{'i}_L=d^i_L$,
$u^{i'}_L=V_{ik}u^{k}_L$, and $\nu^{'i}_L=U_{ik}\nu^k_L$ (down alignment), where $V$ and $U$ are the Cabbibo-Kobayashi-Maskawa (CKM)  and the Ponte\--corvo-Maki-Nakagawa-Sakata (PMNS) mixing matrices, respectively. This leads  to the following interactions of the $S_1$ and $\widetilde{R}_2$ representations with neutrinos and quarks
\begin{align}
\mathcal{L}_{\widetilde{R}_2}&\supset\widetilde{R}_2^{2/3}V_{ik}\widetilde{Y}^{\overline{LR}}_{2kj}\bar{u}^i_L \nu^j_R\nonumber\\
&+\widetilde{R}_2^{-1/3}\left(U_{kj}\widetilde{Y}^{RL}_{2ik}\bar{d}^i_R \nu^j_L
+\widetilde{Y}^{\overline{LR}}_{2ij}\bar{d}^i_L \nu^j_R\right)+{\rm H.c.},
\end{align}
and
\begin{align}
\mathcal{L}_{S_1}&\supset S_1^{1/3}\left(U_{kj}Y_{1\,ik}^{LL}\bar{d}_L^{Ci}\nu_L^j + Y^{\overline{RR}}_{1\,ij}\bar{d}_R^{Ci}\nu_R^j \right)+{\rm H.c.}
\end{align}

For completeness we present in Table \ref{LQcoupneut_} the LQ interactions to a neutrino-quark pair of all the scalar and vector LQs arising from all the representations of Table \ref{LQrep}.

\begin{table*}[!htb]
\caption{Interactions of the scalar and vector LQs arising from the representations of Table \ref{LQrep} with a quark and left- and right-handed neutrinos.   \label{LQcoupneut_}}
\begin{center}
\begin{tabular}{cc}
\hline\hline
Representation&Neutrino interactions\\
\hline
$S_3$&$-(Y_3^{LL}U)_{ij}\bar{d}_L^{Ci}\nu_L^j S_3^{1/3}+\sqrt{2}(V^TY^{LL}_3U)_{ij}\bar{u}_L^{Ci}\nu_L^j S_3^{-2/3}+{\rm H.c.}$\\
$R_2$& $(Y^{RL}_{2}U)_{ij} \bar{u}^{i}_R \nu^j_L {R}_2^{2/3}+{\rm H.c.}$\\
$\widetilde{R}_2$&
$(\widetilde{Y}^{RL}_{2}U)_{ij}\bar{d}^i_R \nu^j_L \widetilde{R}_2^{-1/3}+\widetilde{Y}^{\overline{LR}}_{2\,ij}\bar{d}^i_L \nu^j_R\widetilde{R}_2^{-1/3}
+(V\widetilde{Y}^{\overline{LR}}_{2})_{ij} \bar{u}^i_L \nu^j_R \widetilde{R}_2^{2/3}
+{\rm H.c.}$\\
$S_1$&$(Y_1^{LL}U)_{ij}\bar{d}_L^{Ci}\nu_L^j S_1^{1/3}+ Y^{\overline{RR}}_{1\,ij}\bar{d}_R^{Ci}\nu_R^j S_1^{1/3}+{\rm H.c.}$\\
$\bar{S}_1$&$\overline{Y}^{\overline{RR}}_{1\,ij}\bar{u}_R^{Ci}\nu_R^j \bar{S}_1^{-2/3}+{\rm H.c.}$\\
$U_3$&$(V X^{LL}_{3}U)_{ij} \bar{u}^i_L\gamma^\mu \nu^j_L U^{2/3}_{3\,\mu}+
\sqrt{2}(X^{LL}_{3}U)_{ij}\bar{d}^i_L\gamma^\mu \nu^j_L U^{-1/3}_{3\,\mu}
 +{\rm H.c.}$\\
$V_2$&$-(X_{2}^{RL}U)_{ij}\bar{d}_R^{Ci}\gamma^\mu\nu_L^j V_{2\,\mu}^{1/3}+{\rm H.c.}$\\
$\widetilde{V}_2$&$-(\widetilde{X}_{2}^{RL}U)_{ij}\bar{u}_R^{Ci}\gamma^\mu\nu_L^j\widetilde{V}^{-2/3}_{2\,\mu}+ (V^T\widetilde{X}_{2}^{\overline{LR}})_{ij}\bar{u}_L^{Ci}\gamma^\mu\nu_R^j \widetilde{V}^{-2/3}_{2\,\mu}-\widetilde{X}^{\overline{LR}}_{2\,ij}\bar{d}_L^{Ci}\gamma^\mu\nu_R^j \widetilde{V}^{1/3}_{2\,\mu}+{\rm H.c.},$\\
$U_1$&$(VX^{LL}_{1}U)_{ij}\bar{u}^{i}_L\gamma^\mu \nu^{j}_L U^{2/3}_{1\,\mu}+
X^{\overline{RR}}_{1\,ij}\bar{u}^{i}_R\gamma^\mu \nu^{j}_R  U^{2/3}_{1\,\mu}$\\
$\widetilde{U}_1$&$X^{\overline{RR}}_{1\,ij}\bar{d}_R^{i}\gamma^\mu\nu_R^j \widetilde{U}_{1\,\mu}^{-1/3}+{\rm H.c.}$\\
\hline\hline
\end{tabular}
\end{center}
\end{table*}

For the purposes of our calculation, we will consider a generic interaction of a scalar LQ $\Phi^k$ of electric charge $Q_k=k$, in units of $|e|$,  to  quarks and Dirac neutrinos of the form
\begin{align}
\label{genSLag}
\mathcal{L}^{\Phi^k q^i\nu^\alpha}=&\bar{q}^i\left(\zeta^0_{L\,i\alpha} P_L+\zeta^0_{R\,i\alpha} P_R\right)\nu^\alpha \Phi^{k}+
{\rm H.c.},
\end{align}
where the quark $q^i$ is of up type (down type) for the scalar LQ $\Phi^{2/3}$ $(\Phi^{-1/3})$.  As usual $P_{L,R}$ are the left- and right-handed chirality projectors and the LQ coupling constants $\zeta^0_{L\,i\alpha}$ and $\zeta^0_{R\,i\alpha}$, where the superscripts stands for the LQ spin and the  column (row) index is denoted by a Latin  (Greek letter) and it is associated with the quark (lepton) family, will be assumed as complex and can be extracted from Table \ref{LQcoupneut_}.  Notice however  that the  scalar LQ representation $S_1$  gives rise to  interactions with charge conjugate quark fields of the form
\begin{align}
\label{MajoranaScaLQ}
\mathcal{L}^{\Phi^k q^{C,i}\nu^\alpha}&=\bar{q}^{C, i}\left(\tilde{\zeta}^0_{L\,i\alpha} P_L+\tilde{\zeta}^0_{R\,i\alpha} P_R\right)\nu^\alpha \Phi^{k\,*}+
{\rm H.c.},
\end{align}
where again the quark $q^i$ is of up type (down type) for the scalar LQ $\Phi^{2/3}$ $(\Phi^{-1/3})$.
We will discuss below that our results for the NEPs obtained from the interaction \eqref{genSLag} can be used {\it mutatis mutandi} to obtain the contributions of the scalar LQs that obey the interaction  \eqref{MajoranaScaLQ}.

\subsubsection{Electromagnetic interactions of scalar LQs}
As far as the couplings of the scalar doublets to the photon are concerned, they  can be straightforwardly obtained from the  kinetic  Lagrangian, which for a scalar LQ multiplet $\Phi$ can be generically written as
\begin{equation}
\label{LQKinetic}
{\cal L}_{\text{kin}}^{\text{LQ}}=\frac{1}{2}(D_\mu \Phi)^\dagger D^\mu \Phi,
\end{equation}
where the $SU(2)\times U(1)$ covariant derivative is given by
\begin{equation}
\label{CovDer}
D_\mu=\partial_\mu+ig I^l W^l_\mu+ig' Y_\Phi B_\mu,
\end{equation}
where $l$ runs from 1 to 3 and $I^k$ are matrices in the $SU(2)$ representation of $\Phi$: $I^l=0$ for $SU(2)$ singlets, $I^l=\tau^l$ ($l=1,2,3$), with $\tau^l$ being the Pauli matrices, for  $SU(2)$ scalar LQ doublets, and $\left(I^l\right)_{ij}=-i\epsilon_{ijl}$ ($i,j,l=1,2,3$)  for  $SU(2)$ scalar LQ triplets. As usual $B_\mu$ and $W^i_\mu$ are the abelian and nonabelian gauge fields.

After rotating to the mass eigenstates, we obtain the couplings of the LQ multiplet components $\Phi^k$  to the photon, which  can be written as
\begin{equation}
\mathcal{L}^{A\Phi^k\Phi^{k\dagger}}=ieQ_k  A^\mu\left(\Phi^{k\dagger}\partial_\mu \Phi^k-\Phi^k\partial_\mu \Phi^{k\dagger} \right).
\end{equation}

\begin{figure*}
\centering
\includegraphics[width=14cm]{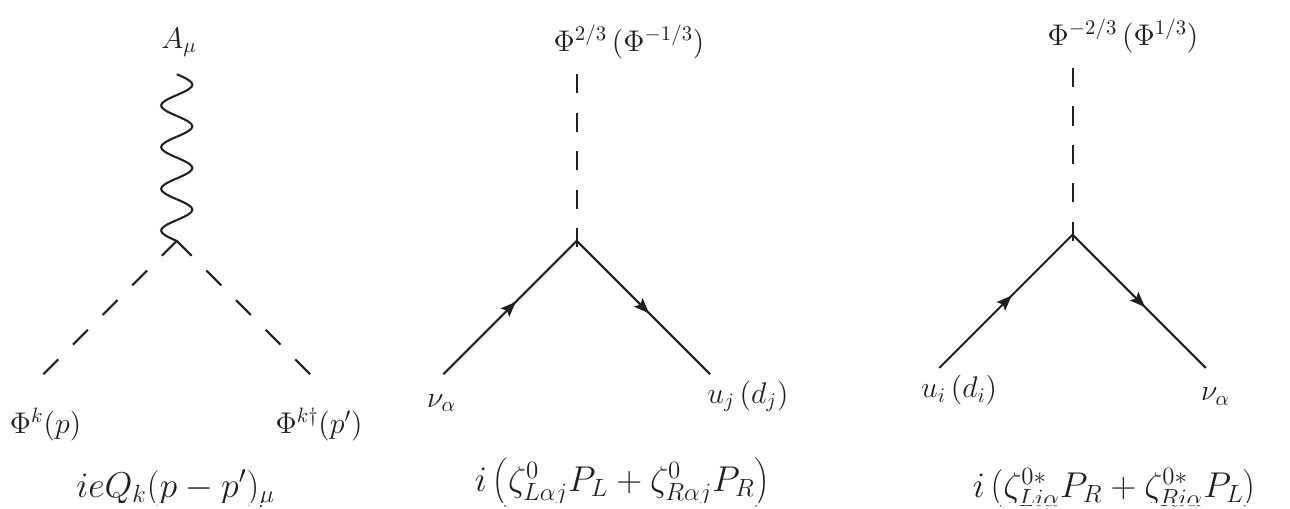}
\caption{Generic Feynman rules for the interactions of a scalar LQ $\Phi^k$ necessary for the calculation of  NEPs. Here $k$ stands for the LQ electric charge in units of $|e|$. The usual Feynman rules for the particle propagators and the SM interactions are not shown.\label{FeynmanRulesScaLQ}}
\end{figure*}

Apart from the usual SM Feynman rules, the remaining ones necessary for our calculation can be obtained from the above Lagrangians and  are presented in  Fig. \ref{FeynmanRulesScaLQ}. 

\subsection{Vector LQ interactions}
\subsubsection{$U_1$ and $\widetilde{V}_2$ neutrino couplings}
As for the vector $U_1$ and $\widetilde{V}_2$   representations,  their interactions with the left- and right-handed neutrinos follow from the current-sector Lagrangian and can be written as follows
\begin{align}
\mathcal{L}_{U_1}&\supset
\left(X^{LL}_{1\,ij}\bar{Q}^{'i}_L\gamma^\mu L^{'j}_L+X^{\overline{RR}}_{1\,ij}\bar{u}^{'i}_R\gamma^\mu \nu^{'j}_R
\right)U_{1\mu}+
\text{H.c}.,
\end{align}
and 
\begin{align}
\mathcal{L}_{\widetilde{V}_2}&\supset 
\widetilde{X}^{RL}_{2\,ij}\bar{u}^{'C\,i}_R\gamma^\mu \widetilde{V}_{2\,\mu}i\tau_2 L^{'j}_L +\widetilde{X}^{\overline{LR}}_{2\,ij}\bar{Q}^{'C\,i}_L\gamma^\mu i\tau_2\widetilde{V}_{2\,\mu} \nu^{'j}_R\nonumber\\&+
\text{H.c}.,
\end{align}
where again the coupling constants can be taken as complex quantities  in general, but they are fixed to gauge coupling constants in the case of gauge LQs. Note that the $\widetilde{V}_2$ representation induces diquark couplings, so the couplings of the $\widetilde{V}_2^{2/3}$ and $\widetilde{V}_2^{-1/3}$ components can be severely constrained from proton decay  unless an ad hoc symmetry is invoked to achieve proton stability, We will only consider this representation to present  the most general calculation of vector LQ contributions to NEPs.

After rotating to the mass eigenstates we obtain the following interactions of vector LQs with left- and right-handed neutrinos
\begin{align}
\mathcal{L}_{U_1}&\supset U^{2/3}_{1\,\mu}\left(V_{im}U_{kj}X^{LL}_{1\,mk}\bar{u}^{i}_L\gamma^\mu \nu^{j}_L+
X^{\overline{RR}}_{1\,ij}\bar{u}^{i}_R\gamma^\mu \nu^{j}_R\right)\nonumber\\&
+\text{H.c.},
\end{align}
and
\begin{align}
\mathcal{L}_{\widetilde{V}_2}&\supset\widetilde{V}^{-2/3}_{2\,\mu}\left(-U_{ik}\widetilde{X}^{RL}_{2\,kj}\bar{u}_R^{C\,i}\gamma^\mu\nu_L^j+ V^T_{ik}\widetilde{X}^{\overline{LR}}_{2\,kj}\bar{u}_L^{C\,i}\gamma^\mu\nu_R^j\right)\nonumber\\& -\widetilde{V}^{1/3}_{2\,\mu}\widetilde{X}^{\overline{LR}}_{2\,ij}\bar{d}_L^{Ci}\gamma^\mu\nu_R^j +{\rm H.c.},
\end{align}

Again, we will consider a generic interaction of a vector LQ $V^k$ of electric charge $Q_k=k$, in units of $|e|$, to charge $Q_k$ quarks and neutrinos of the form
\begin{align}
\label{genVLag}
\mathcal{L}^{V^k_\mu q^i\nu^\alpha}=&V^{k}_\mu
\bar{q}^i\gamma^\mu\left(\zeta^1_{L\,i\alpha} P_L+\zeta^1_{R\,i\alpha} P_R\right)\nu^\alpha 
+{\rm H.c.},
\end{align}
where the quark $q^i$ is of up type (down type) for the vector LQ $V^{2/3}$ $(V^{-1/3})$. The LQ coupling constants $\zeta^1_{L\,i\alpha}$ can be obtained from Table \ref{LQcoupneut_} for all the vector LQ representations of Table \ref{LQrep}.
As in the case of scalar LQs, we will see below that the results for NEPs obtained from \eqref{genVLag}  will also allows one to obtain the results for the contribution arising from the $\widetilde{V}_2$ representation, which yields an interaction of the form
\begin{align}
\label{MajoranaVecLQ}
\mathcal{L}^{V^k_\mu q^{C,i}\nu^\alpha}=&
\bar{q}^{C,i}\gamma^\mu\left(\tilde{\zeta}^1_{L\,i\alpha} P_L+\tilde{\zeta}^1_{R\,i\alpha} P_R\right)\nu^\alpha V^{k\dagger}_\mu+
{\rm H.c.}
\end{align}

\subsubsection{Electromagnetic interactions of vector LQs}

As far as the LQ electromagnetic couplings, for simplicity we will consider below vector LQs that are arise from  a gauge theory spontaneously broken.
Following our model-independent approach we consider that once the gauge group of the ultraviolet (UV) completion has been broken into the SM gauge group, the gauge LQ interactions with the SM gauge bosons are given by the most general renormalizable $SU(2)\times U(1)$ invariant Lagrangian for a gauge LQ $V_\mu$ \cite{Dorsner:2016wpm,Gabrielli:2015hua,Montano:2005gs}
\begin{equation}
\label{gaugeLQ}
{\cal L}^{V_\mu}=-\frac{1}{2}V^\dagger_{\mu\nu}V^{\mu\nu}-ig' V^{\dagger \mu}B_{\mu\nu}V^{\nu}-ig V^{\dagger \mu}\vec{W}_{\mu\nu}V^{\nu},
\end{equation}
where $V_{\mu\nu}=D_\mu V_{\nu}-D_\nu V_{\mu}$, with the $SU(2)_L\times U(1)_Y$ covariant derivative given in Eq. \eqref{CovDer}, also $\vec{W}_{\mu\nu}=I^k W_{\mu\nu}^k$, with the matrices $I^k$ being defined above.

Finally, after the gauge symmetry has broken to the $U(1)_{\rm em}$ group and all the gauge fields have been rotated to their mass eigenstates, we arrive at the following interaction Lagrangian of a pair of gauge LQs $V^k_\mu$ to the photon \cite{Ferrara:1992yc}
\begin{equation}
{\cal L}^{AV^{k\dagger} V^k}=-\frac{1}{2}V^{k\,\dagger}_{\mu\nu}V^{k\,\mu\nu}-i Q_ke V^{k\,\dagger}_\mu F^{\mu\nu}V^k_\nu,
\end{equation}
where $F_{\mu\nu}$ is the electromagnetic  strength tensor and
\begin{align}
V^k_{\mu\nu}&\equiv D^{\rm em}_\mu V^k_{\nu}-D^{\rm em}_\nu V^k_{\mu}
\nonumber\\&=\left(\partial_\mu+iQ_k eA_\mu \right) V^k_{\nu}-\left(\partial_\nu+iQ_keA_\nu \right) V^k_{\mu}.\end{align}

\begin{figure*}[!hbt]
\centering
\includegraphics[width=14cm]{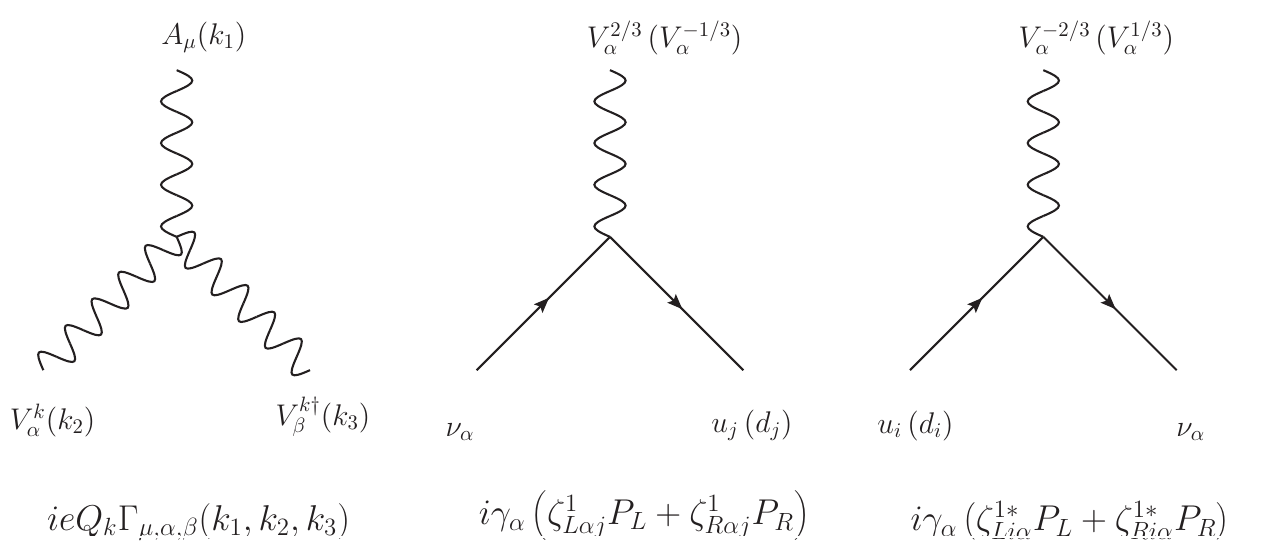}
\caption{Generic Feynman rules for a gauge LQ $V^k_\mu$ in the unitary gauge necessary for the calculation of the neutrino static dipole moments. Here $k$ stands for the LQ electric charge in units of $|e|$ and the $AV^{k\dagger} V^k$ vertex function is $\Gamma^{\rm UG}_{\mu,\alpha,\beta}(k_1,k_2,k_3)=(k_3-k_2)_\mu g_{\alpha\beta}+(k_1-k_3)_\alpha g_{\beta\mu}+(k_2-k_1)_\beta g_{\mu\alpha}$.
 The usual Feynman rules for the particle propagators and SM interactions are not shown.\label{FeynmanRulesVecLQ}}
\end{figure*}

The generic Feynman rules in the unitary gauge for a gauge LQ are presented in Fig. \ref{FeynmanRulesVecLQ}, though they are only useful to calculate the static neutrino dipole moments, which are gauge-independent quantities.

However, to obtain the effective NCR we need to make some assumptions about the UV completion of the LQ model since the calculation must be performed via the BFM to obtain a gauge-independent result.
As far as  the LQ interactions with the SM gauge bosons are concerned,  as shown in Ref. \cite{Montano:2005gs},
where  the BFM formalism was used to obtain the Feynman rules  for the new singly and doubly charged gauge bosons arising  in an  $SU(3)_L\times U(1)_X$ model, once the extended gauge symmetry is spontaneously broken  into $SU(2)_L\times U(1)_Y$, the vertex functions for the trilinear and quartic gauge boson couplings share the same Lorentz structure, which stems from the fact that they all obey $SU(2)_L\times U(1)_Y$ symmetry. Even more,  the couplings of any charged gauge boson and its associated pseudo-Goldstone boson must obey electromagnetic gauge invariance. Thus, the vertex function  for the coupling of a photon with a pair of charged gauge bosons $AVV^\dagger$ is identical for any charged gauge boson except for the electric charge factor, and the same is true for the vertex function of the  $AG_VG_V^\dagger$ coupling, where $G_V$ is the pseudo-Goldstone boson associated with the charged gauge boson (note that in the BFM there is no mixed tree-level $AV G_V^\dagger$ coupling).  Thus,  the Feynman-rules for the  $AVV^\dagger$ and $AG_V G_V^\dagger$ couplings, with $V$ a gauge LQ and $G_V$ its associated pseudo-Goldstone boson, must be analogue to those presented in \cite{Denner:1994xt} and \cite{Montano:2005gs}  for singly and doubly charged gauge bosons after replacing the LQ electric charge.

\begin{figure*}[!hbt]
\centering
\includegraphics[width=14cm]{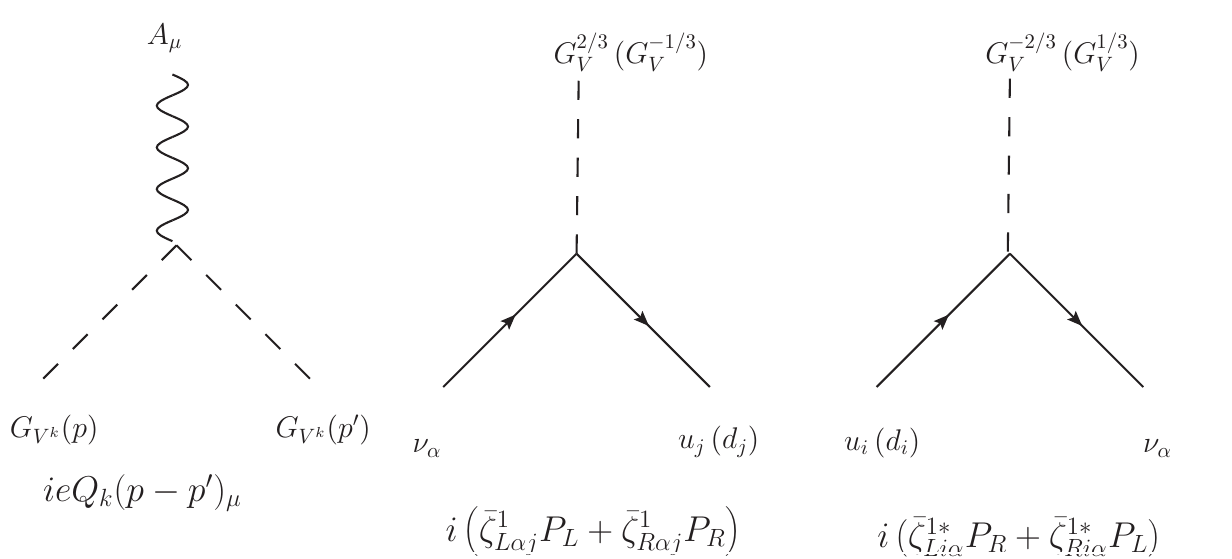}
\caption{Feynman rules in the Feynman-'t Hooft gauge of the BFM, which together with those of Fig. \ref{FeynmanRulesVecLQ} are necessary for the calculation of the contribution of a gauge LQ to the effective NCR.  Note that in this gauge  the $AVV^\dagger$ vertex function of  Fig. \ref{FeynmanRulesVecLQ} is given by  $\Gamma^{\rm BFM}_{\mu,\alpha,\beta}(k_1,k_2,k_3)=(k_3-k_2)_\mu g_{\alpha\beta}+(k_1-k_3-k2)_\alpha g_{\beta\mu}+(k_2-k_1+k3)_\beta g_{\mu\alpha}$. In addition, the left- and right handed coupling constants of the gauge LQ and its associated pseudo-Goldstone boson to the fermions must be fixed to the gauge constants. Thus, inspired in the model of Ref. \cite{DiLuzio:2017vat,DiLuzio:2018zxy} we assume a simple renormalizable gauge LQ model where the coupling constants of a gauge LQ and its associated pseudo-Goldstone bosons to a neutrino-quark pair  are given as in   Table \ref{CouConsGauLQ}.
\label{FeynmanRulesVecLQBFM}}
\end{figure*}

 Nevertheless, for the interactions of the LQ and its pseudo-Goldstone boson with the SM fermions we do need to know more details of the UV completion of the LQ model since the coupling constants of Eq. \eqref{genVLag} are fixed to the gauge coupling constants. To obtain an estimate of the magnitude of the LQ contributions to the effective NCR we will assume a gauge LQ with left-handed coupling inspired in the model of \cite{DiLuzio:2017vat,DiLuzio:2018zxy}. The corresponding Feynman rules are presented in Fig. \ref{FeynmanRulesVecLQBFM}, where we also include the Feynman rules for the coupling of a pseudo-Goldstone boson to a neutrino-quark pair, with the coupling constants being given in Table \ref{CouConsGauLQ}.

\begin{table}[!htb]
\caption{Left- and right-handed coupling constants for the interactions of a gauge LQ and its associated pseudo-Goldstone boson to the a neutrino-quark pair, inspired in the model of Ref. \cite{DiLuzio:2017vat}. Here $g_4$ stands for a gauge coupling constant. \label{CouConsGauLQ}}
\begin{tabular}{ccc}
\hline
\hline
Vertex&Left-handed couplings&Right-handed couplings\\
\hline
\hline
$V^k \bar{q}^i\nu^\alpha$&$\zeta^1_{L\,i\alpha}=\dfrac{g_4}{\sqrt{2}}\beta_{i\alpha}$&$\zeta^1_{R\,i\alpha}=0$\\
$G_V^{k} \bar{q}^i\nu^\alpha$&$\bar\zeta^1_{L\,i\alpha}=\dfrac{g_4}{\sqrt{2}}\dfrac{m_{q_i}}{m_{\rm LQ}}\beta_{i\alpha}$&$\bar\zeta^1_{R\,i\alpha}=\dfrac{g_4}{\sqrt{2}}\dfrac{m_{\nu_\alpha}}{m_{\rm LQ}}\beta_{i\alpha}$\\
\hline
\hline
\end{tabular}
\end{table}

\section{LQ contribution to NEP\lowercase{s}}
\label{Calculation}

\begin{figure*}[!hbt]
\centering
\includegraphics[width=14cm]{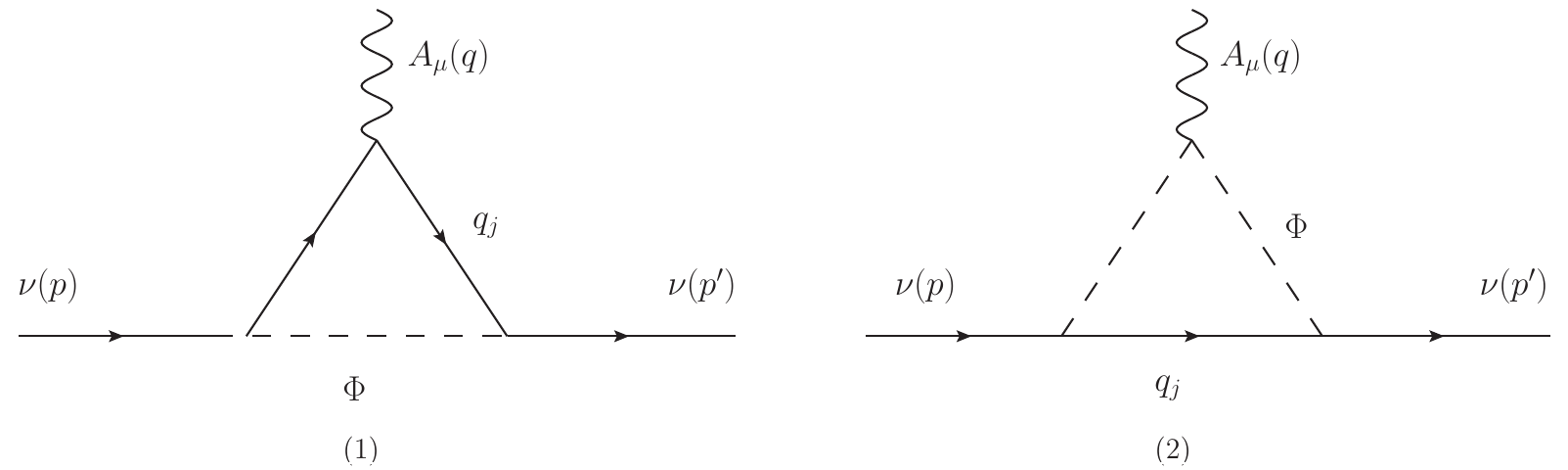}
\caption{Feynman diagrams that contribute to the neutrino dipole moments and the effective NCR at the one-loop level in models with scalar LQs. \label{FeynmanDiagramsScaLQ}}
\end{figure*}

\begin{figure*}[!hbt]
\centering
\includegraphics[width=15cm]{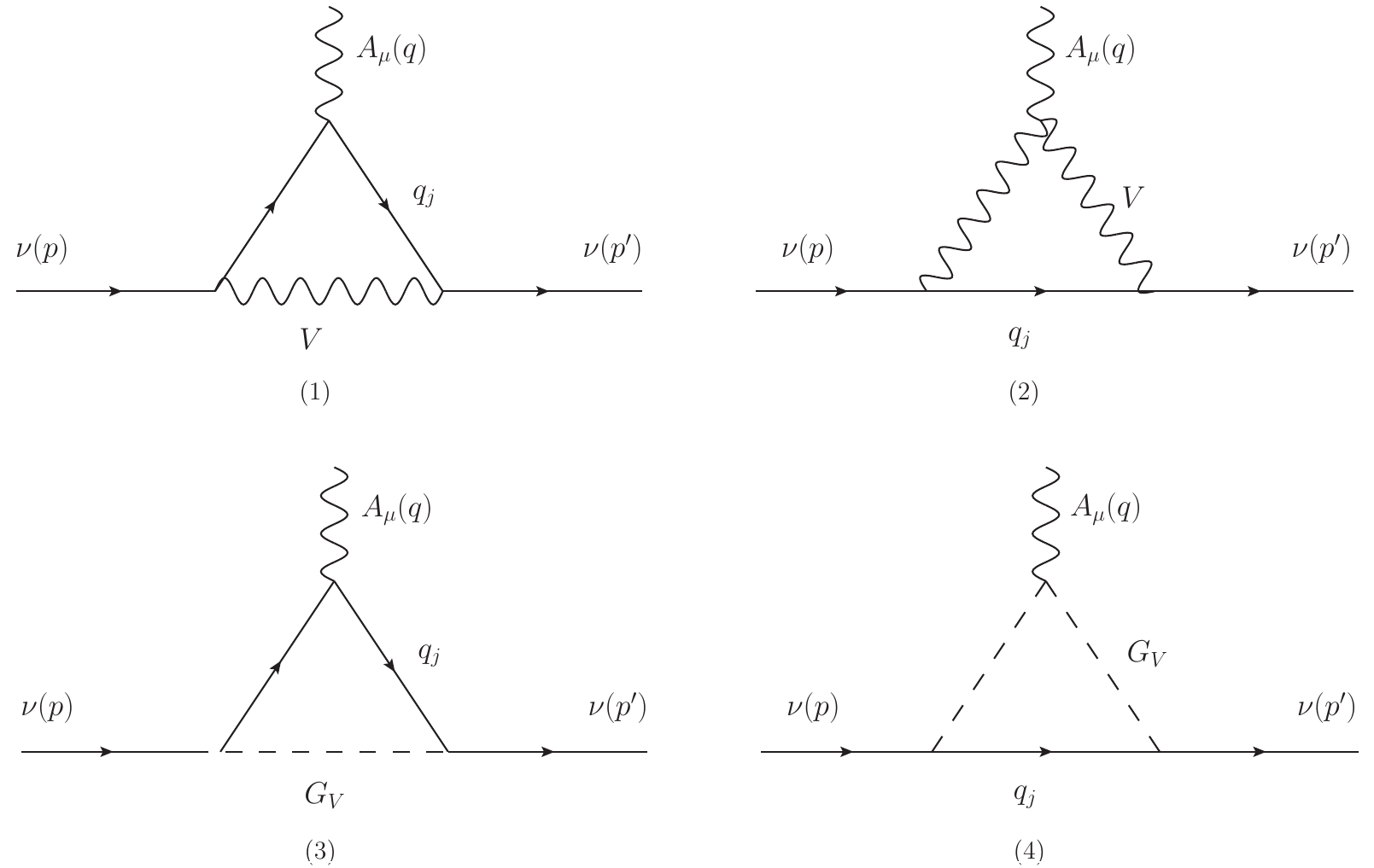}
\caption{Feynman diagrams for the contribution to  NEPs from gauge LQ $V^k_\mu$. The upper row diagrams can be calculated in the unitary gauge to obtain the contributions to the neutrino MDM and EDM, whereas the four diagrams must be calculated in the Feynman-'t Hooft gauge of the BFM to obtain the contribution to the effective NCR. Here $G_V$ stands for the  pseudo-Goldstone boson associated to  $V$. \label{FeynmanDiagramsVLQs}}
\end{figure*}

We now turn to present our calculation of the LQ contributions to the NEPs, which at the one-loop level are induced by the Feynman diagrams of Fig. \ref{FeynmanDiagramsScaLQ} for scalar LQs
and Fig. \ref{FeynmanDiagramsVLQs} for vector LQs. As already mentioned,    the $\gamma-Z$ self-energy diagrams  does not contribute to the effective NCR, as discussed in \cite{Papavassiliou:2003rx}.

The loop amplitudes were worked out by  Feynman-parameter integration with the help of the FeynCalc package \cite{Mertig:1990an,Shtabovenko:2020gxv} to perform the Dirac algebra. An independent evaluation was done by means of the Passarino-Veltman reduction scheme via the FeynCalc  and  Package-X routines \cite{Patel:2015tea}, which allowed us to make a cross-check. We first obtained results for nonvanishing $q^2$ and afterwards a careful procedure was applied to obtain the LQ contributions to the NEPs in the limit of $q^2=0$.

We first present the most general form of the contribution of a spin-$s$ LQ to the NEPs, and detailed results for scalar and gauge LQs will be given below.

The contribution to the MDM of neutrino $\nu_\alpha$ can be written in Bohr magneton units as follows
\begin{align}
\label{nuMDM}
\mu_{\nu_\alpha}^s&=\frac{N_cQ_{\rm LQ}m_e}{16\pi^2m_{\rm LQ}}\sum_{i}\Bigg(\left(\frac{m_{\nu_\alpha}}{m_{\rm LQ}}\right)\left(\left| \zeta^{s}_{L\,i\alpha}\right|{}^2+\left|\zeta^{s}_{R\,i\alpha}\right| {}^2\right)\nonumber\\&\times f^{s}(x_{q_i})
+\left(\frac{m_{q_i}}{m_{\rm LQ}}\right) {\rm Re}\left(\zeta^{s}_{L\,i\alpha} \zeta^{{s}*}_{R\,i\alpha}\right)g^{s}(x_{q_i})\Bigg)\;\mu_B,
\end{align}
where the superscript $s$ stands for the LQ spin, whereas $Q_{\rm LQ}$ and $m_{LQ}$ denote its electric charge and mass. We also define $x_{q_i}=m_{q_i}^2/m_{\rm LQ}^2$, with $q_i$ being the virtual quark. The $f^{s}(x)$ and $g^{s}(x)$ functions can be written as
\begin{align}
\label{ftot}
f^{s}(x)&=\sum_j f^{s}_j(x),
\end{align}
and
\begin{align}
\label{gtot}
g^{s}(x)&=\sum_j g^{s}_j(x),
\end{align}
where the $f^s_j(x)$ and $g^s_j(x)$ functions stand for the contribution of each  Feynman diagram, which will be presented below in approximate and full form.  The LQ coupling constants $\zeta^{s}_{L\,i\alpha}$ and $\zeta^{s}_{R\,i\alpha}$ can be extracted from Table \ref{LQcoupneut_} for the scalar and vector LQs arising from the representations of Table \ref{LQrep}.

As for the contribution of a spin-$s$ LQ to the neutrino electric dipole moment, it requires complex LQ couplings and can be written in terms of the $g^s(x)$ functions as follows
\begin{align}
\label{nuEDM}
d_{\nu_\alpha}^s&=-\frac{e N_cQ_{\rm LQ}}{32\pi^2m_{\rm LQ}}\sum_{i}\left(\frac{m_{q_i}}{m_{\rm LQ}}\right) {\rm Im}\left(\zeta^s_{L\,i\alpha} \zeta^{s*}_{R\,i\alpha}\right) g^s(x_{q_i}).
\end{align}

As far as   the effective NCR is concerned, following Refs. \cite{Bernabeu:2000hf,Bernabeu:2002pd}, we obtain the contributions of scalar and vector LQs to the $\hat \Gamma^\mu_{A\bar{\nu}\nu}$ vertex for nonvanishing $q^2$, from which the dimensionless effective form factor $\hat F_1^V(q^2)$ can be extracted as the coefficient of $ie\gamma^\mu(1-\gamma^5)$: it is given in terms of the dimensionful form factor $\hat F_{\nu_\alpha}(q^2)$ as follows
\begin{equation}
\hat F^{V}(q^2)= q^2 {\hat F}_{\nu_\alpha}(q^2).
\end{equation}
The effective NCR is thus given by $\langle r^2 \rangle_{\nu_\alpha}=6 \hat F_{\nu_\alpha}(0)$. While the scalar LQ contribution to $\hat F^{V}(q^2)$ is gauge-independent and  was calculated straightforwardly,  the vector LQ contribution was obtained via the BFM in the Feynman-'t Hooft gauge, as described more detailed below.

The LQ contribution to the effective NCR  can be written as
\begin{align}
\label{nuCR}
\langle r^2 \rangle_{\nu_\alpha}^s&=\frac{N_cQ_{\rm LQ}}{16\pi^2m_{\rm LQ}^2}\sum_{i}\Bigg(\left(\left| \zeta^s_{L\,i\alpha}\right|{}^2+\left|\zeta^s_{R\,i\alpha}\right| {}^2\right)\hat f^s(x_{q_i})\nonumber\\&
+\left(\frac{m_{q_i} m_{\nu_\alpha}}{m_{\rm LQ}^2}\right) {\rm Re}\left(\zeta^s_{L\,i\alpha} \zeta^{s*}_{R\,i\alpha}\right)\hat g^s(x_{q_i})\Bigg),
\end{align}
where the $f^s(x)$ and $g^s(x)$ obey  relationships similar to Eqs. \eqref{ftot} and \eqref{gtot}. 

We have obtained  results for the $r^s_k(x)$ functions ($r=f,\,g,\,\hat{f},\,\hat{g}$) for nonzero neutrino mass in terms of both Feynman-parameter integrals and Passarino-Veltman scalar functions, which are presented in   \ref{AnalyRes} and can be useful to obtain the NEPs of hypothetical heavy Dirac neutrinos. From such results, approximate expressions were obtained to leading order in the neutrino mass, which provide a good estimate for the NEPs of light neutrinos, and are presented below.

\subsection{Scalar LQ contribution for light neutrinos}
The scalar LQ contributions to the NEPs  are clearly gauge-independent and yield ultraviolet finite MDM and EDM. As for the contribution  to the neutrino charge form factor $\hat F_1^V(q^2)$, its derivative is finite at $q^2=0$ and so is  the effective NCR.

The contributions of a scalar LQ to the neutrino MDM are given trough
the  $f^0_i(x)$ and $g^0_i(x)$ functions of Eqs. \eqref{nuMDM} and can be approximated for light neutrinos as
\begin{align}
\label{ffunc1}
f^0_1(x)&=-\frac{1}{6\left(x-1\right){}^4} \left(\left(\left(x-6\right) x+3\right) x+6 x \log \left(x\right)\right.\nonumber\\&+\left.2\right),\\
\label{ffunc2}
f^0_2(x)&=-\frac{1}{6 \left(x-1\right){}^4} \left(\left(x-1\right) \left(x \left(2 x+5\right)-1\right)\right.\nonumber\\&-\left.6 x^2 \log \left(x\right)\right),
\end{align}
and
\begin{align}
\label{gfunc1}
g^0_1(x)&=-\frac{1}{\left(x-1\right){}^3} \left(x^2-4 x+2 \log \left(x\right)+3\right),\\
\label{gfunc2}
g^0_2(x)&=-\frac{1}{\left(x-1\right){}^3} \left(x^2-2 x \log \left(x\right)-1\right).
\end{align}
The latter functions are also useful to compute the contribution of a scalar LQ to the neutrino EDM of Eq. \eqref{nuEDM}.

As far as the contributions of a scalar LQ to the effective NCR,  the  $\hat f^0(x)$ and $\hat g^0(x)$ functions of Eq. \eqref{nuCR} are given as follows  for light neutrinos
\begin{align}
\label{fhatfunc1}
\hat f^0_1(x)&=\frac{1}{12 \left(x-1\right){}^4}\left(6 \left(2-3 x\right) \log \left(x\right)
\right.\nonumber\\&-\left.\left(x-1\right) \left(x \left(7 x-29\right)+16\right)\right),\\
\label{fhatfunc2}
\hat f^0_2(x)&=-
\frac{1}{12\left(x-1\right){}^4}
\left(\left(x-1\right) \left(x \left(11 x-7\right)+2\right)\right.\nonumber\\&-\left.6 x^3 \log \left(x\right)\right),
\end{align}
and
\begin{align}
\label{ghatfunc1}
\hat g^0_1(x)&=-\frac{1}{3 \left(x-1\right){}^5}\left(\left(x-1\right) \left(\left(x-8\right) x-17\right)
\right.\nonumber\\&+\left.6 \left(3 x+1\right) \log \left(x\right)\right),\\
\label{ghatfunc2}
\hat g^0_2(x)&=\frac{1}{3  \left(x-1\right){}^5} \left(\left(\left(9-17 x\right) x+9\right) x
\right.\nonumber\\&+\left.6 \left(x+3\right) x^2 \log \left(x\right)-1\right).
\end{align}

\subsection{Vector LQ contribution}
We first calculate the contribution of a vector LQ to the static neutrino MDM and EDM, which are gauge independent and can thus be straightforwardly computed  in the unitary gauge, where the contributing Feynman diagrams  are the ones shown in the upper row of Fig. \ref{FeynmanDiagramsVLQs}.

The  results for the $f^1_k(x)$ and $g^1_k(x)$ functions for light neutrinos read as follows

\begin{align}
f^1_1(x)&=\frac{1}{6 \left(x-1\right){}^4} \left(18 x^2 \log \left(x\right)\right.\nonumber\\&-\left.\left(x-1\right) \left(x \left(x \left(5 x-9\right)+30\right)-8\right)\right),\\
f^1_2(x)&=-\frac{1}{6 \left(x_q-1\right){}^4} \left(18 x^3 \log \left(x\right)\right.\nonumber\\&+\left.\left(x-1\right) \left(x \left(x \left(4 x-45\right)+33\right)-10\right)\right),\\
g^1_1(x)&=\frac{1}{\left(x-1\right){}^3} \left(x^3+3 x-6 x \log \left(x\right)-4\right),\\
g^1_2(x)&=\frac{1}{\left(x-1\right){}^3} \left(\left(x-1\right) \left(\left(x-11\right) x+4\right)
\right.\nonumber\\&+\left.6 x^2 \log \left(x\right)\right).
\end{align}

As far as the calculation of the effective NCR is concerned, the gauge LQ contribution to the $\hat F^V(q^2)$ form factor must be obtained via the BFM in the Feynman-'t Hooft gauge. Apart from the Feynman diagrams with an internal gauge LQ,  there are two additional Feynman diagrams where the gauge LQ is replaced by its associated pseudo-Goldstone boson, as shown in the lower row of Fig. \ref{FeynmanDiagramsVLQs}. Nevertheless, the latter contributions are suppressed by a factor of $m_{q_i}/m_{\rm LQ}$ and even for the top quark will give a subdominant contribution.

According to the coupling constants assumed in Table \ref{CouConsGauLQ}, the gauge LQ contribution to the effective NCR
can be written as in Eq. \eqref{nuCR} but with vanishing right-handed couplings, namely, $\zeta^1_{R\,i\alpha}=0$. It means that there are only contributions to the effective NCR via the  $\hat f_j(x)$ functions, which are given by
\begin{align}
\hat f^1_1(x)&=\frac{1}{6 \left(x_q-1\right){}^4}\left(\left(x-1\right) \left(x \left(25 x-29\right)-2\right)
\right.\nonumber\\&-\left.
6 \left(6 x^2-9 x+2\right) \log \left(x\right)\right),\\
\hat f^1_2(x)&=\frac{1}{6 \left(x-1\right){}^4}\left(6 x^2 \left(5 x-6\right) \log \left(x\right)
\right.\nonumber\\&-\left.
\left(x-1\right) \left(x \left(43 x-65\right)+16\right)\right),
\end{align}
\begin{align}
\hat f^1_3(x)&=\frac{x}{12 \left(x-1\right){}^4}\left(6 \left(2-3 x\right) \log \left(x\right)
\right.\nonumber\\&-\left.\left(x-1\right) \left(x \left(7 x-29\right)+16\right)\right),
\end{align}
\begin{align}
\hat f^1_4(x)&=\frac{x}{12 \left(x-1\right){}^4} \left(6 x^3 \log \left(x\right)
\right.\nonumber\\&-\left.\left(x-1\right) \left(x \left(11 x-7\right)+2\right)\right),
\end{align}
where again the subscript  stands for the contribution of each Feynman diagram of Fig. \ref{FeynmanDiagramsVLQs}.

Note that for a very heavy LQ, namely, $x_{q_i}\ll 1$ the effective NCR for a gauge LQ with the couplings of Table \ref{CouConsGauLQ}   can be approximated as
\begin{align}
\label{nuVCRapr}
\langle r^2 \rangle_{\nu_\alpha}^1&\simeq\frac{N_cQ_{\rm LQ}}{16\pi^2m_{\rm LQ}^2}\sum_{i}\left| \zeta^1_{L\,i\alpha}\right|{}^2
\left(3-2\log\left(x_{q_i}\right)\right),
\end{align}
which reduces to the SM result of Eq. \eqref{SMNCRRes} after the replacements: $\left| \zeta^1_{L\,i\alpha}\right|{}^2\to \dfrac{g^2}{2}=2\sqrt{2}m_W^2 G_F$,  $m_{\rm LQ}\to m_W$, $m_{q_i}\to m_e$, $Q_{\rm LQ}\to 1$, and $N_c\to 1$.

It is also worth noting that the  LQs arising from the $S_1$ and $\widetilde{V}_2$ representations involve Feynman rules of  Majorana type, which requires special treatment \cite{Denner:1992me,Denner:1992vza}. Nevertheless, the corresponding contributions to NEPs are identical to those obtained for the $\widetilde{R}_2$ and $U_1$ representations. Those the above results are valid for all the LQ representations of Table \ref{LQrep}.  A similar situation was discussed in \cite{Bolanos:2019dso,Bolanos-Carrera:2022iug} for the contribution of scalar LQs to the   $f_i\to f_j\gamma$ and $t\to c\gamma\gamma$ decays.

\section{Numerical Analysis}
\label{NumAnal}
We now turn to the numerical analysis, for which we need to discuss the current constraints on the LQ mass and couplings consistent with experimental limits on high precision observables and direct searches at the LHC. A LQ model meant to solve the LFUV anomalies in $B$-meson decays and the $(g-2)_\mu$ anomaly may require several ingredients, such as extra LQ representations,  new symmetries, and fine tuning. Note also that such anomalies, if confirmed by future measurements, could also be explained by another mechanism of the UV completion of the LQ model and not necessarily by the LQs. In our analysis we will consider instead a simple LQ model with only one LQ representation, which will allow to asses the potential LQ contributions to NEPs.

\subsection{Constraints on the masses of scalar and vector LQs}
\label{LQmasscons}

The most recent constraints on the  masses of both scalar and vector LQs have been obtained from direct searches at the CERN LHC by the CMS and ATLAS collaborations using the data for proton collisions at $\sqrt{s}=13$ TeV \cite{CMS:2018ncu,CMS:2018lab,CMS:2018oaj,CMS:2018svy,ATLAS:2020dsk,ATLAS:2020xov,ATLAS:2021jyv,CMS:2022zks,ATLAS:2022wcu,CMS:2022nty,CMS:2020wzx} via single  and double production of LQs  decaying into a quark and a lepton, namely,  $pp\to   \Phi \bar\ell\to q\ell\bar\ell$ and $pp\to  \Phi^\dagger \Phi \to q\bar{q}\ell\bar\ell, q\bar q \nu\bar \nu$. The constraints obtained this way are rather model dependent since it is usually assumed that LQs only couple to fermions of one generation and  have a dominant decay channel,  though LQs that can couple  to fermions of distinct generations have also been considered recently \cite{CMS:2018oaj,ATLAS:2020dsk,ATLAS:2020xov}. In addition, since the oblique parameters strongly constrain the mass splitting of a $SU(2)$ multiplet \cite{Crivellin:2020ukd}, LQs of the same $SU(2)$ representation are assumed mass degenerate. 

\subsubsection{Charge $2/3$ LQs} 

The ATLAS and CMS collaborations \cite{ATLAS:2021jyv,ATLAS:2023uox,CMS:2020wzx}  set a lower  bound on the mass of a third-generation scalar LQ  
ranging  between 900 and 1450 GeV for a branching ratio of the $ b\tau$ decay channel ranging from 10 to 100 percent. On the other hand,  for a vector LQ in the Yang-Mills scenario the lower mass bound  goes from 1400 to about 1950 GeV for a branching ratio of the $ b\tau$ decay channel ranging from 10 to 100 percent \cite{ATLAS:2023uox,CMS:2020wzx}, whereas  in the so called minimal coupling scenario the respective lower mass bounds are about 300 GeV less stringent \cite{ATLAS:2023uox,CMS:2020wzx}.

The constraints are more severe when it is imposed the condition that a solution to the LFUV anomalies in $B$-meson decays is imposed. In this scenario, the ATLAS collaboration has searched for pair produced scalar and vector LQs decaying into a quark of the third generation accompanied by a lepton of the second and first generations \cite{ATLAS:2022wcu,CMS:2022nty}: for the mass of a charge $2/3$ scalar LQ decaying into the $b\mu$ ($be$) pair  a lower bound  of about 1440 (1460) GeV was found. Along the same line, the CMS collaboration has  reported \cite{CMS:2022zks} a search for third-generation charge $2/3$ scalar and vector LQs decaying into a bottom quark plus a $\tau$ lepton  via single and pair LQ production: when the LQ coupling to the $b\tau$ pair is of the order of unity,  the lower bound on the mass of a scalar (vector) LQ is about 1250 GeV (1530 GeV), whereas for a LQ coupling of the order of 2.5 the  corresponding lower mass limit is about 1370 GeV (1960 GeV) for a scalar (vector) LQ.

\subsubsection{Charge $-1/3$ LQs}

ATLAS has searched for this type of LQs \cite{ATLAS:2023kek}  via pair production and its decay into a $\tau$ lepton plus a charm  or a lighter quark. A charge $-1/3$ LQ with mass below 1.3 TeV is excluded as long as the branching fraction for the $\tau c$ decay channel is of one hundred percent. The CMS collaboration also searched for scalar and vector LQs decaying into a neutrino-quark pair and found the bounds of 1110 GeV for a scalar LQ and $1475$-$1810$ GeV for a vector LQ \cite{CMS:2018qqq}.  

In the scenario where a solution to the LFUV anomalies is required, the ATLAS collaboration has searched for this type of LQ via the decay into $t\mu$ ($t e$) and set the lower bound  of about 1380  GeV (1370 GeV); again, the bounds on the mass of vector  LQs are more stringent\cite{ATLAS:2022wcu}:  for a charge $-1/3$ vector LQ decaying into $t\mu$ ($t e$), the lower bound  is 1980 GeV (1900 GeV) in the Yang-Mills scenario, but such a bound relaxes to 1710 GeV (1620 GeV) in the minimal coupling scenario.

We present a summary of the above constraints in Table \ref{LQmassbounds}. For our numerical analysis we will consider LQs with masses from $1.2$ to $2$ TeVs.

\begin{table*}[!hbt]
\caption{Current lower bounds on the mass of scalar and vector LQs from direct searches by the ATLAS and CMS collaborations \cite{CMS:2018ncu,CMS:2018lab,CMS:2018oaj,CMS:2018svy,ATLAS:2020dsk,ATLAS:2020xov,ATLAS:2021jyv,CMS:2022zks,ATLAS:2022wcu,CMS:2022nty,CMS:2020wzx}. In scenario I the condition of a solution to the LFUV anomalies in $B$-meson decays is not imposed, whereas in scenario II such a solution is required indeed.  See the text and References therein for the assumptions made to obtain such limits.\label{LQmassbounds}}
\begin{center}
\begin{tabular}{cccc}
\hline
\hline
LQ charge&Scenario&Scalar LQ mass (GeV)& Vector LQ mass (GeV)\\
\hline
\hline
$2/3$&I&900-1450&1400-1950\\
$2/3$&II&1250-1450&1460-1960\\
$-1/3$&I&1300&$1475$-$1810$\\
$-1/3$&II&1370-1380&1620-1980\\
\hline
\hline
\end{tabular}
\end{center}
\end{table*}

\subsection{Realistic scenarios for the LQ coupling constants}

In order to obtain a realistic estimation of the NEPs we must consider  scenarios for the LQ couplings to fermion bilinears consistent with the most up-to-date constraints on high-precision experimental observables. 
We will focus on those scenarios that could allow for the largest values for the LQ coupling constants as they can give the largest contributions to NEPs. 

LQ particles can have dangerous effects on electroweak precision observable quantities, which may yield strong constraints on the LQ masses and couplings. As far as LQ couplings to fermions are concerned,  LQ couplings  to the fermions of the first-generation are strongly constrained by low-energy processes, such as atomic parity violation  \cite{Langacker:1990jf,Leurer:1993em,Davidson:1993qk}, universality in leptonic pion decays \cite{Shanker:1982nd,Davidson:1993qk,Leurer:1993em}, $\mu-e$ conversion \cite{Shanker:1981mj},
flavor changing Kaon decays \cite{Shanker:1981mj,Valencia:1994cj},  as well as $K^0- \overline{K}^0$ and $D^0-\overline{D}^0$ mixings \cite{Shanker:1982nd,Davidson:1993qk,Leurer:1993ap}. We will thus assume that an extra symmetry forbids the LQ couplings to the first generation of fermions while still allowing nonzero couplings to the fermions of the second and third generations, as has been customary done in the literature.  

An analysis of the constraints on the coupling constants of all the LQ representations of Table \ref{LQrep} is beyond the scope of this work. Thus, for our analysis we will focus only on the scalar $\widetilde{R}_2$ and vector $U_1$ representations as they are two of the LQ representations that can yield the largest possible values of NEPs as will be shown below.

\subsubsection{The $\widetilde{R}_2$ representation}

This LQ representation can give a solution to the anomalies in $B$-meson decays, but  additional representations are required to explain the discrepancy in the muon $g-2$ \cite{Chen:2022hle}: it turns out that apart from the charge $-1/3$ LQ that only couples to quarks and neutrinos, this representation yields a charge $2/3$ chiral LQ that can couple to down quarks and charged leptons, thereby inducing a  contribution to the muon $g-2$ lacking of an enhancing  chirality-flipping term. 

We will content ourselves with discussing the constraints on the LQ couplings in a simple model in which only one $\widetilde{R}_2$ LQ representation is introduced with both left- and right handed couplings to neutrinos.  Constraints on these couplings can be obtained from   the $B_s-\overline{B}_s$ mass difference  and the decays $b\to s\mu^-\mu^+$,  $B\to K\nu\nu$, $\tau\to\mu\gamma$,  $\tau\to \mu\phi$ as well as the semileptonic meson decays such $B\to \tau\nu$, $D_s\to \tau\nu$, etc. The analytical expressions for the contribution of the $\widetilde{R}_2^{2/3}$ LQ  to these observable quantities can be found in \cite{Becirevic:2016oho}, but for easy reference they are presented 
in \ref{lowenerobser}. In our analysis we will assume that the PMNS mixing matrix is approximately diagonal, which means that the $\zeta^0_{L}$ and $\zeta^0_{R}$ couplings from Eq. \eqref{genSLag} can be identified with the Yukawa couplings $\widetilde{Y}^{RL}_{2}$ and $\widetilde{Y}^{\overline{RR}}_{2}$, respectively, of Table \ref{LQcoupneut_}. We will also assume the following ansatz to avoid the strong constraints from parity violation
\begin{equation}
\label{ansatzRt2}
{\zeta}^0_{L}\simeq
\begin{pmatrix}
0&0&0\\
0&\zeta^0_{L\, 2\mu}&\zeta^0_{L\, 2\tau}\\
0&\zeta^0_{L\, 3\mu}&\zeta^0_{L\, 3\tau}
\end{pmatrix},
\end{equation}
with a similar expression for ${\zeta}^0_R$.  
 
\begin{figure*}[!htb]
\centering
\includegraphics[width=12cm]{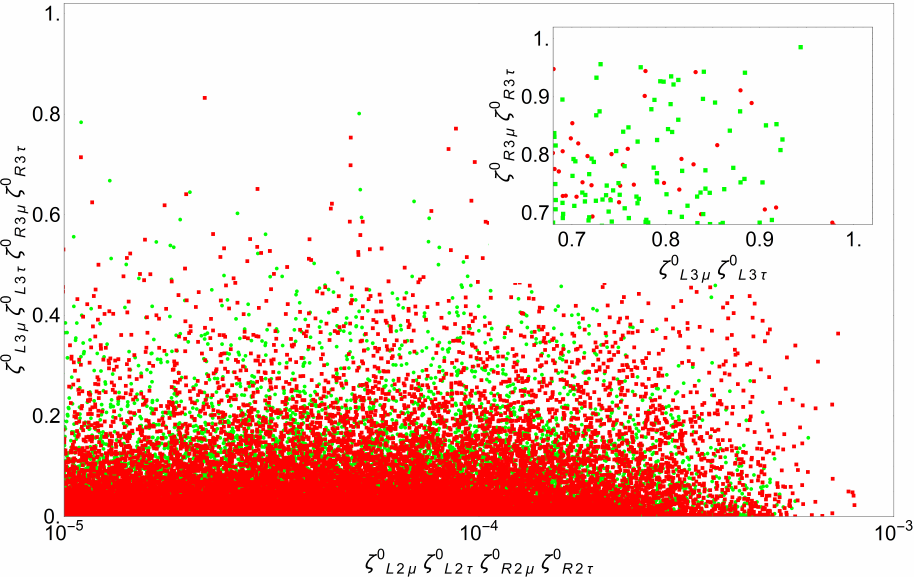}
\caption{Allowed  values of the Yukawa couplings of the $\widetilde{R}_2$ representation in the plane $\zeta^{0}_{L\, 2\mu}\zeta^{0}_{L\, 2\tau} \zeta^{0}_{R\, 2\mu}\zeta^{0}_{R\, 2\tau}$  vs $\zeta^{0}_{L\, 3\mu}\zeta^{0}_{L\, 3\tau} \zeta^{0}_{R\, 3\mu}\zeta^{0}_{R\, 3\tau}$ for  LQ masses of $1.2$ TeVs (green points) and $1.5$ TeVs (red points).  We use the experimental limits on the  $B_s-\overline{B}_s$ mass difference as well as the decays $b\to s\mu^-\mu^+$,   $B\to K\nu\nu$, $\tau\to\mu\gamma$, and $\tau\to \mu\phi$.  The $\zeta^{0}_{L\, 2 \alpha}$ and $\zeta^{0}_{R\, 2 \alpha}$ ($\alpha=\mu,\,\tau$) couplings were taken to be below $0.1$ to allow for large values of the $\zeta^{0}_{L\, 3 \alpha}$ and $\zeta^{0}_{R\, 3 \alpha}$ couplings, for which we impose the perturbativity bound  $\zeta^{0}_{L\, 3 \alpha},\,\zeta^{0}_{R\, 3 \alpha}\le 1$, which is also meant to evade the constraints from direct searches at the LHC.    \label{RtCoupBounds}}
\end{figure*}

Since we are interested in the scenario that can yield the largest estimates for the NEPs, we will  consider small values for the  coupling constants $\zeta^0_{L\, 2\alpha}$ and $\zeta^0_{R\, 2\alpha}$,  below the $0.1$ level, which will allow for  large  values of $\zeta_{L\, 3\alpha}$ and $\zeta_{R\, 3\alpha}$. 
We randomly scan over  sets of values of the nonzero $\zeta^0_{L,R}$  matrix elements   and  found those sets that fulfil  the experimental limits on the $B_s-\overline{B}_s$ mass difference  as well as the decays $b\to s\mu^-\mu^+$ and $B\to K\nu\nu$. We also impose the bounds $\zeta^0_{L\, 3\alpha},\,\zeta^0_{R\, 3\alpha}\le 1$ to evade the constraints from direct searches at the LHC and avoid the breakdown of perturbation theory. Although we  include in our analysis other observables such as the lepton flavor violating (LFV) tau decays $\tau\to\mu\gamma$ and $\tau\to \mu\phi$, as well as semileptonic meson decays, they do not yield useful constraints for small $\zeta^0_{L\, 2\alpha}$ and $\zeta^0_{R\, 2\alpha}$.  We show in Fig. \ref{RtCoupBounds}  the allowed values of the LQ couplings in the
$\zeta^{0}_{L\, 2\mu}\zeta^{0}_{L\, 2\tau} \zeta^{0}_{R\, 2\mu}\zeta^{0}_{R\, 2\tau}$ vs $\zeta^{0}_{L\, 3\mu}\zeta^{0}_{L\, 3\tau} \zeta^{0}_{R\, 3\mu}\zeta^{0}_{R\, 3\tau}$ for LQ masses of 1.2 TeV and 1.5 TeVs. We can conclude that there are allowed values of the product  $\zeta^{0}_{L\, 3\mu}\zeta^{0}_{L\, 3\tau} \zeta^{0}_{R\, 3\mu}\zeta^{0}_{R\, 3\tau}$ close to the unity, which means that  there are regions of the parameter space where  $\zeta^{0}_{L\, 3\alpha}$ and $\zeta^{0}_{R\, 3\alpha}$ can be  close to one simultaneously, as can be observed in the zoomed area.  Note also that such values are also consistent with  the model-independent bounds on the LQ couplings from direct searches at the LHC compiled in \cite{Schmaltz:2018nls}.

\subsubsection{The $U_1$ representation} 
This representation is appealing as it  gives no tree-level contribution to  observables  such as the $B_s-\overline{B}_s$ mass difference and the very constrained decay $B\to K\nu\bar{\nu}$, though loop contributions  can be nonegligible. Nevertheless,  finding constraints on the $U_1$ couplings  can be troublesome as its contribution to radiative corrections can be plagued with quadratic divergences, thereby requiring  a specific UV completion   to tackle this issue.  As already mentioned, a few relatively simple UV completions into which the $U_1$ representation can be embedded  have been discussed in Refs. \cite{Bordone:2017bld,Buttazzo:2017ixm,Calibbi:2017qbu,DiLuzio:2017vat,Blanke:2018sro,Barbieri:2017tuq,DiLuzio:2018zxy}, where the authors try to address some dangerous effects on low energy observables that can push the LQ mass well beyond the TeV scale such as in the original Pati-Salam model. Those models contain new particles, apart from the $U_1$ LQ, that contribute to low-energy observables, thus extra assumptions are required to avoid tension with experimental measurements. 

\begin{figure*}
\centering
\includegraphics[width=12cm]{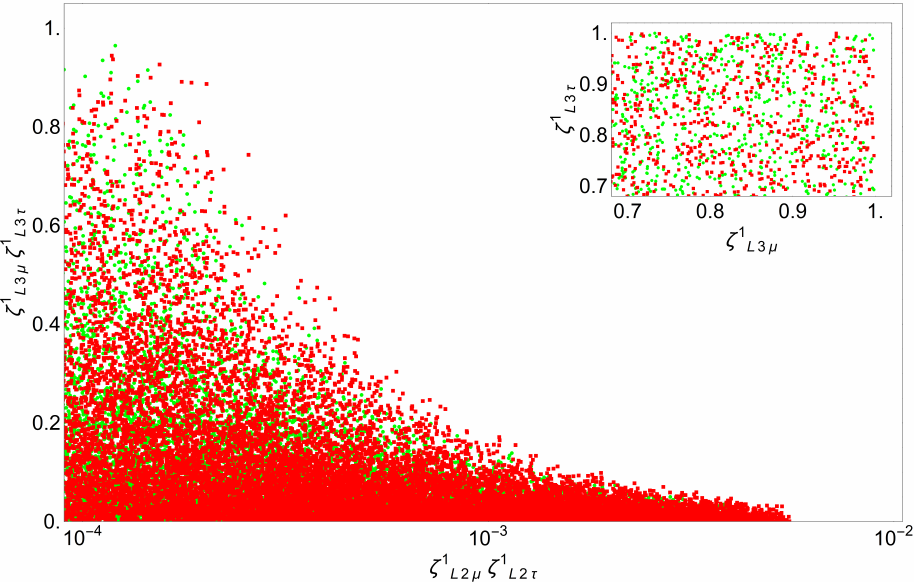}
\caption{Allowed  values for the Yukawa couplings of the $U_1$ representation  in the plane $\zeta^{1}_{L\, 2\mu}\zeta^{1}_{L\, 2\tau}$  vs $\zeta^{1}_{L\, 3\mu}\zeta^{1}_{L\, 3\tau} $ for a LQ masses of $1.5$ TeVs (green points) and $1.7$ TeVs (red points)  from the experimental limits on  the decays $b\to s\mu^-\mu^+$, $\tau\to \mu\phi$.  The $\zeta^{1}_{L\, 2 \alpha}$  ($\alpha=\mu,\,\tau$) couplings were taken to be below $0.1$ to find the largest allowed values of the $\zeta^{1}_{L\, 3 \alpha}$ couplings, for which we impose the perturbativity bound  $\zeta^{1}_{L\, 3 \alpha}\le 1$. We consider that either there are no right-handed neutrinos or the corresponding coupling constants are too small.    \label{U1CoupBounds}}
\end{figure*}

We will consider a model with a lone vector $U_1$ LQ that only couples to the fermions of the second and third generations in the scenario where either  there are no right-handed neutrinos or the corresponding coupling constants are negligible, such as in the model of Ref. \cite{DiLuzio:2018zxy}. For the LQ couplings to fermions we will use $\zeta^1_{L\,i\alpha}=g_4\beta_{i\alpha}/\sqrt{2}$ and $\zeta^1_{L\,i\alpha}=0$  with $g_4\sim 3$ \cite{DiLuzio:2018zxy}.
Again, we consider  $\zeta^1_{L\, 2\mu},\,\zeta^1_{L\, 2\tau}\le 0.1$  to let the  $\zeta^1_{L\, 3\alpha}$ couplings  reach its largest allowed values, below the perturbativity bound of $1$. We then scan over  sets of random $\zeta^1_{L\, i\alpha}$ values and  select those sets  consistent with  the experimental limits on  the decays $b\to s\mu^-\mu^+$, $B\to K\tau^-\tau^+$, $B^+\to K^+\tau^\pm\mu^\mp$, and $\tau\to \mu\phi$. Other $b$-hadron decays such as $B\to \tau\nu$, $B_c\to \tau \nu$, $B\to \tau^-\tau^+$, and $B_s\to \tau^-\mu^+$   yield no useful constraints.  The corresponding expressions are given in  \ref{lowenerobser} for easy reference.  We show in Fig. \ref{U1CoupBounds}  the allowed values for the LQ couplings in the
$\zeta^{1}_{L\, 2\mu}\zeta^{1}_{L\, 2\tau}$ vs $\zeta^{1}_{L\, 3\mu}\zeta^{1}_{L\, 3\tau}$ plane for LQ masses of 1.5 TeV and 1.7 TeVs.   We observe that  the product  $\zeta^{1}_{L\, 3\mu}\zeta^{1}_{L\, 3\tau}$ can reach values of the order of $O(1)$ for $\zeta^{1}_{L\, 2\mu}\zeta^{1}_{L\, 2\tau}$ below  $10^{-4}$, which means that  the $\zeta^{1}_{L\, 2\mu}$ and $\zeta^{1}_{L\, 2\tau}$ couplings are allowed to take on  values of the order of $O(1)$ simultaneously, as can be observed in the zoomed area.  As in the case of the $\widetilde{R}_2$ representations, the allowed values for the $U_1$ coupling constants are consistent with the model-independent bounds on the LQ couplings from direct searches at the LHC compiled in Ref. \cite{Schmaltz:2018nls}.

Note also that in the models of \cite{Bordone:2017bld,Buttazzo:2017ixm,Calibbi:2017qbu,DiLuzio:2017vat,Blanke:2018sro,Barbieri:2017tuq,DiLuzio:2018zxy}  there are extra particles such as new gauge and scalar bosons as well as additional fermions, which can give new contributions to low energy observables. For instance in the model of \cite{DiLuzio:2017vat} there are two extra neutral gauge bosons $g'$ and $Z'$, as well as new scalar bosons and fermions, whereas the model presented in \cite{Bordone:2017bld} also includes right-handed neutrinos.  In those works, a complete treatment of the constraints on the parameter space from experimental data was performed  with the purpose of addressing the LFUV anomalies in $B$-meson decays. In general, one can assume that the  LQ couplings to the quarks of the second family are negligible, which allows for  large couplings  to the quarks of the third generation.   

Therefore, in the most promising scenario we can assume values for the couplings of scalar and vector LQs to quarks and fermions of the order of $O(1)$, which would lead to the best scenario for the magnitude of the LQ contribution to the NEPs of the tau and muon neutrino, whereas those of the electron neutrino would be strongly suppressed as the couplings to the first generation fermions would only arise from quark mixing.

\subsection{Behavior of the LQ contribution to NEPs}

The general behavior of the LQ contribution to NEPs can be inferred from Eqs. \eqref{nuMDM}, \eqref{nuEDM}, and \eqref{nuCR}, which show the presence of chirality-flipping $(\rm CF)$  and chirality-conserving $(\rm CC)$ terms: the MDM and effective NCR  have the following structure
\begin{align}
\label{defmdm}
\mu^s_{\nu_\alpha}&= a^s_{\rm CC} (|\zeta^s_{L3\alpha}|^2+|\zeta^s_{R3\alpha}|^2)+a^s_{\rm CF} {\rm Re}(\zeta^s_{L3\alpha}\zeta^{s*}_{R3\alpha}),
\end{align}
and 
\begin{align}
\label{defncr}
\langle r^2 \rangle^s_{\nu_\alpha}&=b^s_{\rm CC} (|\zeta^s_{L3\alpha}|^2+|\zeta^s_{R3\alpha}|^2)+b^s_{\rm CF} {\rm Re}(\zeta^s_{L3\alpha}\zeta^{s*}_{R3\alpha}),
\end{align}
whereas for the  EDM  we have
\begin{align}
\label{defedm}
d^s_{\nu_\alpha}=\widetilde{a}^s_{\rm CF}\, {\rm Im}(\zeta^s_{L3\alpha}\zeta^{s*}_{R3\alpha}).
\end{align}
Note that  $a^s_{\rm CF}\sim \widetilde{a}^s_{\rm CF}\sim b^s_{\rm CC} \sim m_{q_j}$, whereas $a^s_{\rm CC}\sim b^s_{\rm CF}\sim m_{\nu_\alpha}$. 
Thus, the most promising scenario for sizeable LQ contributions to  both dipole moments is that in which the LQ couples to both left- and right-handed neutrinos, whereas in the absence of such couplings the EDM would vanish and the MDM  would be proportional to the neutrino mass, thereby being negligible small for light neutrinos.

We can conclude that the largest contributions to the electromagnetic dipole moments of light neutrinos could arise from the vector representations  $U_1$ and $\widetilde{V}_2$,  as well as the scalar representations  $\widetilde R_2$ and $S_1$, which are the only ones that can have couplings to both left- and right-handed neutrinos. However, the  contributions from such vector representations are expected to be larger that those of such scalar representations:  the vector LQs have charge $2/3$ and its largest contribution would arise from the loop with the top  quark, whereas  the scalar LQs have charge $-1/3$ and  its dominant contribution would arise from the loop with the much lighter bottom quark. 
As far as the effective NCR is concerned, it   would be nonvanishing even  in absence of LQ couplings to right-handed neutrinos as it receives its dominant contribution from  the $b^s_{\rm CC}$ term, which is proportional to the quark mass. Thus, the LQ contribution to the effective NCR would not be  sensitive to the mass of the virtual quark nor the neutrino mass.

We consider  the scenarios posed by the  scalar and vector LQ representations of Table \ref{LQrep} assuming the presence of a LQ with a mass of 1.5 TeV that can couple to the third-generation quarks and the three SM neutrino flavors. We show in Table \ref{NEPestimates} the estimates for the LQ contributions to the NEPs of a light neutrino with a mass of a few  eVs. For the numerical evaluation we used the approximate expressions for light neutrinos presented above. Since the scalar LQ representations $\widetilde{R}_2$ and $S_1$, along with the vector representations $U_1$ and $\widetilde{V}_2$, are the only ones that yield a LQ that can have simultaneous couplings to both left- and right-handed neutrinos, they would give the largest contributions to the neutrino MDM, which in the best scenario could be of the order of $10^{-10}$-$10^{-9}$. Such representations  could also allow for an EDM  of the order of $10^{-19}$-$10^{-20}$ ecm provided that there are complex LQ couplings. All other LQ representations cannot couple to right-handed neutrinos, so  their contribution to the EDM would vanish, whereas their  contributions to the MDM would be proportional to the neutrino mass, thereby being of the order of $10^{-20}$ $\mu_B$ for $m_{\nu_\alpha}$ in the eV scale.
As far as the effective NCR is concerned, it is insensitive to the neutrino mass and to whether or not the LQs couple to right-handed neutrinos, so all the  LQ representations could yield a contribution of the order of  $10^{-35}$ cm$^2$, regardless of the LQ charge and the accompanying quark. 

Note however that the above estimates must take into account the fact that the coupling constants would not be flavor blind, as observed in our analysis of the bounds on the LQ coupling constants of the $\widetilde{R}_2$ and $U_1$ representations. Thus, an extra suppression is expected for the LQ contribution to the NEPs of each neutrino flavor. For instance, according to the assumptions made to obtain our constraints on the LQ coupling constants, the NEPs of the electron neutrino would be vanishing.

\begin{table*}[!htb]
\caption{Estimate for the contributions of the  LQ representations of Table \ref{LQrep}  to the NEPs of a light Dirac neutrino $\nu_\alpha$ with a mass of 1 eV and a LQ with a mass of  1.5 TeV. It is assumed that a third-generation quark runs into the loop. The coupling constants are those associated with the specific LQ representation and neutrino flavor.   \label{NEPestimates}}
\begin{center}
\begin{tabular}{ccccc}
\hline
\hline
Representation&$\mu_{\nu_\alpha}$ [$\mu_B$]&$d_{\nu_\alpha}$ [ecm]&
$\langle r^2 \rangle_{\nu_\alpha}$ [cm$^2]$\\
\hline
\hline
$S_3$, $R_2$ &$10^{-21}\times |\zeta^0_{L\, 3\alpha}|^2 $&$-$&$10^{-35}\times |\zeta^0_{L\, 3\alpha}|^2$\\
$\widetilde{R}_2$, $S_1$&$10^{-10}\times {\rm Re}(\zeta^0_{L\, 3\alpha} \zeta^{0\,*}_{R\, 3\alpha})$&$10^{-21}\times {\rm Im}(\zeta^0_{L\, 3\alpha} \zeta^{0\,*}_{R\, 3\alpha})$&$10^{-35}\times \left(|\zeta^0_{L\, 3\alpha}|^2+|\zeta^0_{R\, 3\alpha}|^2\right)$\\
$\overline{S}_1$&$10^{-21}\,\times |\zeta^0_{R\, 3\alpha}|^2$&$-$&$10^{-35}\times|\zeta^0_{R\, 3\alpha}|^2$\\
$U_3$, $V_2$&$10^{-20}\,\times |\zeta^1_{L\, 3\alpha}|^2$&$-$&$10^{-35}\times|\zeta^1_{L\, 3\alpha}|^2$\\
 $U_1$, $\widetilde{V}_2$&$10^{-9}\times {\rm Re}(\zeta^1_{L\, 3\alpha} \zeta^{1\,*}_{R\, 3\alpha})$&$10^{-19}\times{\rm Im}(\zeta^1_{L\, 3\alpha} \zeta^{1\,*}_{R\, 3\alpha})$&$10^{-35}\times\left(|\zeta^1_{L\, 3\alpha}|^2+|\zeta^1_{R\, 3\alpha}|^2\right)$\\
$\overline{U}_1$&$10^{-20}\,\times |\zeta^1_{R\, 3\alpha}|^2$&$-$&$10^{-35}\times|\zeta^1_{R\, 3\alpha}|^2$\\
\hline
\hline
\end{tabular}
\end{center}
\end{table*}

\begin{figure*}[!htb]
\centering
\includegraphics[width=15cm]{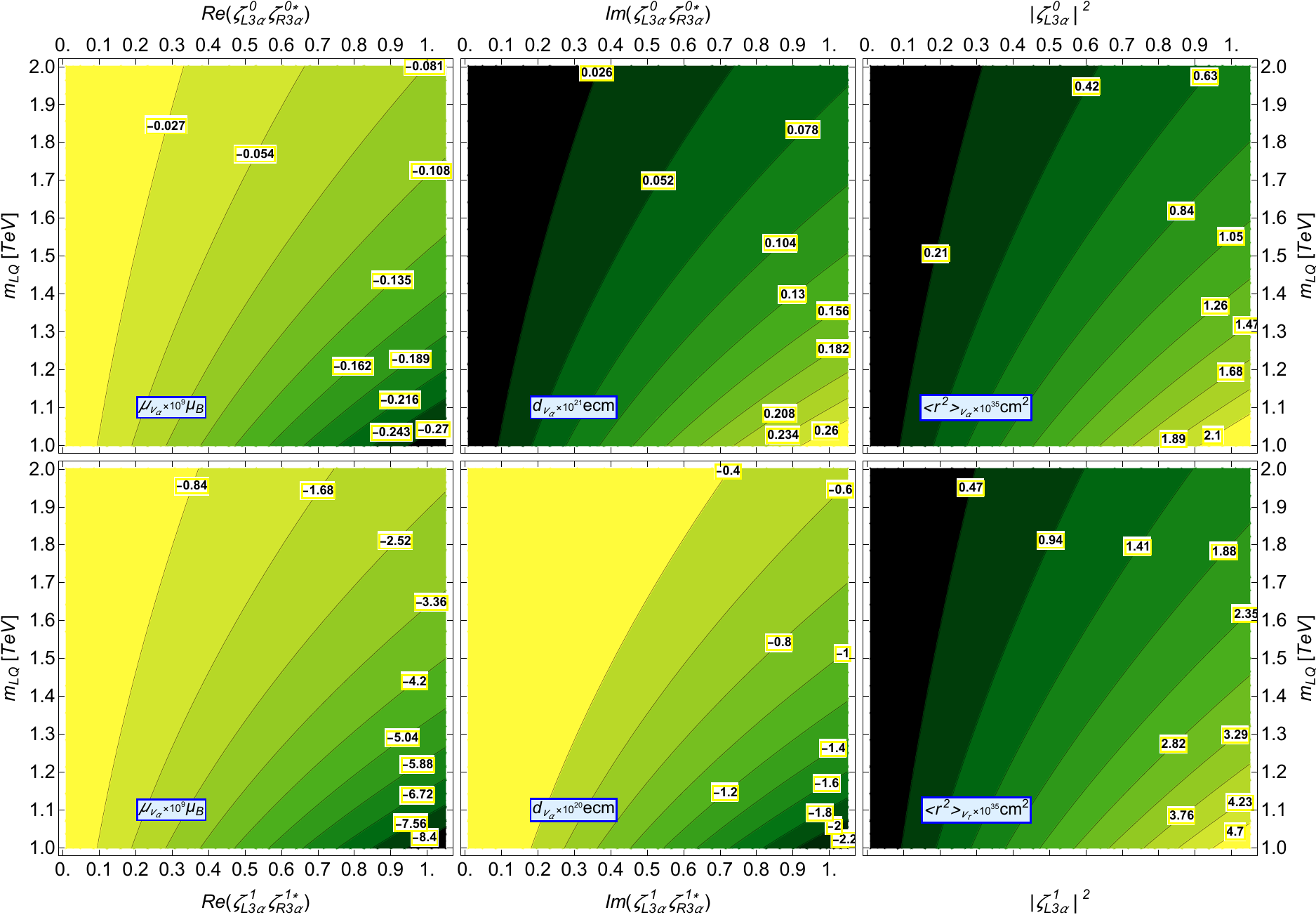}
\caption{Contour plots for the contributions of a scalar LQ of charge $-1/3$ (top row) and a vector LQ of charge $2/3$ (bottom row) to the electromagnetic properties of a light Dirac neutrino $\nu_\alpha$ with a mass of $1$ eV in the plane of the  LQ mass vs the LQ coupling constants. We assume that LQs can couple to both left- and right handed neutrinos. We have neglected the chirality-flipping term for $\mu_{\nu_\alpha}$, whereas for $\langle r^2 \rangle_{\nu_\alpha}$ we assume left-handed neutrinos as it is independent of the presence of right-handed neutrinos.  It is also assumed that the only contribution arises from virtual quarks of the third generation.  \label{tauNEPLQ}}
\end{figure*}

Finally, we consider a charge $-1/3$ scalar LQ and a charge $2/3$ vector LQ,  as the ones arising from the $\widetilde{R}_2$ and $U_1$ representations, in the scenario with right-handed neutrinos. All the features of the behavior of the NEPs described above are best illustrated in Fig.  \ref{tauNEPLQ}, where we show the contours of the LQ contributions to the  NEPs of a neutrino of a mass of $1$ eV as  functions of the LQ mass and the LQ coupling constants.

In closing, we would like to compare our  estimates for the LQ contributions to the NEPs with the experimental constraints of Table \ref{NEPbounds}.
While the LQ contributions to the neutrino MDM are slightly below the experimental constraints, those to the effective NCR are two orders of magnitude smaller. On the other hand, our prediction for the EDM is similar to the existing indirect limits.

It is also worth assessing a possible enhancement to the NEPs in hypothetical LQ models with several LQs: even in the case that the distinct LQ contributions to the NEPs add constructively, an enhancement of several orders of magnitude cannot be expected since all the partial contributions would be about the same order of magnitude in the best scenario. 

\section{Summary}
\label{Summary}

In this work we have presented a calculation of the one-loop contribution of scalar and vector LQ models to the static electromagnetic properties of massive Dirac neutrinos, namely, the magnetic and electric dipole moments as well as the effective NCR defined in \cite{Bernabeu:2002nw}, which is a valid physical observable. We do not make any assumption on the mechanism of neutrino mass generation and consider the effective Lagrangian approach of Buchmuller, Ruckl,  and Wyler for the $SU(3)_c\times SU(2)_L\times U(1)_Y$ scalar and vector LQ representations that have renormalizable couplings to fermion bilinears, including right-handed neutrinos.    Analytical results are presented in terms of both Feynman-parameter integrals and Passarino-Veltman scalar functions in the case of nonzero neutrino mass, which to our knowledge have never been reported  in the literature and could be useful to obtain the LQ contributions to the NEPs of hypothetical heavy neutrinos. From such general results, simple expressions are obtained in the limit of light neutrinos.  It is worth noting that for the vector LQ contribution to the effective NCR we considered a Yang-Mills scenario for gauge LQs and the calculation was performed  via the BFM, which in the Feynman-'t Hooft gauge yields an identical result to that obtained through the  PT. 
For the numerical evaluation we focus on the scalar and vector LQ models that  are renormalizable and do not need  extra symmetries to forbid proton decay at the tree-level, thereby still being phenomenologically viable at the TeV scale, though we also discuss the potential contributions of other LQ representations. 
It is found that the largest contributions to the NEPs of the SM neutrinos may arise from the scalar LQ representation $\widetilde{R}_2$ and  the vector representation $U_1$, which can have simultaneous couplings to both left- and right-handed neutrinos. We then  analyze the current
constraints of the parameter space of such LQ models from experimental data and found that LQ couplings to quarks and neutrinos of the order of $O(1)$ are  still allowed. In such a scenario and for  a LQ with a mass of the order of 1.5 TeV, the $\widetilde{R}_2$ and $U_1$  representations could yield the following contributions to the electromagnetic dipole moments of a light Dirac neutrino with a mass in the eV scale: while the MDM can be of the order of $10^{-9}$ $\mu_B$,  the EDM would be of the order of $10^{-20}-10^{-19}$ ecm  provided that the LQ couplings had  a complex  phase. On the other hand, when there are no LQ couplings to right-handed neutrinos, the EDM would be vanishing and the corresponding LQ contributions to the MDM would be negligible small. As for the  LQ contribution to the effective NCR, it can reach values up to $10^{-35}$ cm$^2$  even in the absence of right-handed neutrinos and regardless of the value of the neutrino mass. Our estimates may have a severe suppression if the LQ couplings to quarks and fermions are below the $O(1)$ level.

\begin{acknowledgements}
We acknowledge support from Conacyt (Mexico). G. Tavares-Velasco also acknowledges partial support from Sistema Nacional de Investigadores (Mexico) and Vicerrector\'ia de Investigaci\'on y Estudios de Posgrado de la Ben\'emerita
Universidad Aut\'onoma de Puebla.
\end{acknowledgements}
\onecolumn
\appendix

\section{Analytical results for the LQ contributions to NEPs}
\label{AnalyRes}
In this appendix we present the  $f^s_k(x)$, $g^s_k(x)$, $\hat f^s_k(x)$, and $\hat g^s_k(x)$ functions ($k=1,2$ and $s=0,1$) of Eqs. \eqref{nuMDM}, \eqref{nuEDM}, and \eqref{nuCR} in terms of both Feynman-parameter integrals and Passarino-Veltman scalar functions for both scalar and vector LQs and nonzero neutrino mass.

\subsection{Feynman parameter results}
In terms of Feynman-parameter integrals, the $f^s_k(x)$, $g^s_k(x)$, $\hat f^s_k(x)$, and $\hat g^s_k(x)$ functions can be cast in the form

\begin{equation}
\label{fsk}
r^s_k(x)=\int_0^1 \frac{R^s_k(x,y)}{((1-y)(x-yx_{\nu_\alpha})+y)^a}dy,
\end{equation}
where  $a=1$ for $r^s_k(x)=f^s_k(x)$ and  $g^s_k(x)$, whereas $a=2$ for $r^s_k(x)=\hat f^s_k(x)$ and $\hat g^s_k(x)$. The respective $R^s_k(x,y)$ functions are given below for both scalar ($s=0$) and vector ($s=1$) LQs.  Note that the dependence of the $r^s_k(x)$ and  $R^s_k(x)$ functions on $x_{\nu_\alpha}$ is not written out explicitly. Here $x_{\nu_\alpha}=m_{\nu_\alpha}^2/m_{\rm LQ}^2$, where $m_{\rm LQ}$ stands for the LQ mass.

\subsubsection{Scalar LQ contribution}
For the contribution of a scalar leptoquark to the neutrino MDM and EDM we obtain
the following expressions for the  $F^s_k(x,y)$ and  $G^s_i(x,y)$ functions:
\begin{align}
F^0_k(x,y)&=\frac{1}{2}(1-y) y\, h_k(y),\\
G^0_k(x,y)&=\frac{1}{3}(1-y)\,h_k(y).
\end{align}
where
\begin{equation}
\label{hfunc}
h_k(y)=\left\{\begin{array}{lcr}
1-y&&k=1,\\
y&&k=2.
\end{array}\right.
\end{equation}

As far as the contribution of a scalar LQ to the effective NCR,  the $\hat F^0_k(x,y)$ and $\hat G^0_k(x,y)$ functions can be written as
\begin{align}
\hat F^0_1(x,y)&=\frac{1}{2}(y-1)^3\left(3 (x+ y^2 {x_{\nu_\alpha}})-2 y \left(x+{x_{\nu_\alpha}}-1\right)\right),\\
\hat F^0_2(x,y)&=\frac{1}{2}y^3 \left((1-y)(x+y {x_{\nu_\alpha}})+y\right),
\end{align}
and
\begin{align}
\hat G^0_1(x,y)&=2 y (y-1)^3,\\
\hat G^0_2(x,y)&=-2 (y-1) y^3.
\end{align}

The integration of the $r^0_k(x)$ functions is straightforward when  $x_{\nu_\alpha}$ is neglected  and yield the results of Eqs. \eqref{ffunc1}, \eqref{ffunc2}, \eqref{gfunc1}, \eqref{gfunc2},  \eqref{fhatfunc1}, \eqref{fhatfunc2},  \eqref{ghatfunc1}, and \eqref{ghatfunc2}.

\subsubsection{Vector LQ contribution}
As far as the contribution of a vector LQ  to the neutrino MDM and EDM,  the  $F^1_k(x,y)$ and $G^1_k(x,y)$ functions are given as follows
\begin{align}
F^1_k(x,y)&=\left(y \left(y \left(x+{x_{\nu_\alpha}}+2\right)-3 x-{x_{\nu_\alpha}}+2\right)+2 x\right)h_k(y),\\
G^1_k(x,y)&=-2\left(y \left((2 y-3) {x_{\nu_\alpha}}+4-x\right)+x+{x_{\nu_\alpha}}\right) h_k(y),
\end{align}
where the $h_k(y)$ function is given in Eq. \eqref{hfunc}.

On the other hand, the $\hat F^1_k(x,y)$ functions associated with the contributions of the Feynman diagrams of Fig. \ref{FeynmanDiagramsVLQs} can be written as

\begin{align}
\hat F^1_1(x,y)&=(y-1) \left(y \left(y \left(-2 y \left(x+{x_{\nu_\alpha}}-1\right)+x+3 y^2 {x_{\nu_\alpha}}+{x_{\nu_\alpha}}+2\right)-2 \left(x+{x_{\nu_\alpha}}-1\right)\right)+3 x\right),
\end{align}
\begin{align}
\hat F_2^1(x,y)&=y^2 \left(y \left(7 x+6({x_{\nu_\alpha}}-1)-y \left(x+(y+5) {x_{\nu_\alpha}}-1\right)+ {x_{\nu_\alpha}}\right)-6 x\right),
\end{align}
\begin{align}
\hat F_3^1(x,y)&=\frac{1}{2}(y-1)^3 \left(3 y^2 x_{\nu_\alpha} \left(x+x_{\nu_\alpha}\right)-2 y \left(x \left(4 x_{\nu_\alpha}-1\right)+x^2+\left(x_{\nu_\alpha}-1\right) x_{\nu_\alpha}\right)+3 x \left(x+x_{\nu_\alpha}\right)\right),
\end{align}
and
\begin{align}
\hat F_4^1(x,y)&=\frac{y^3}{2} \left(-y^2 x_{\nu_\alpha} \left(x+x_{\nu_\alpha}\right)+y \left(4 x x_{\nu_\alpha}-x^2+x+x_{\nu_\alpha}^2+x_{\nu_\alpha}\right)+x \left(x-3 x_{\nu_\alpha}\right)\right).
\end{align}
We note that for the gauge LQ couplings of Table \ref{CouConsGauLQ} there are no contributions to Eq. \eqref{nuCR} of the $\hat G^1_k(x,y)$ functions.

\subsection{Passarino-Veltman results}
For completeness we also present results for the $r^s_k(x)$  functions  in terms of two-point Passarino-Veltman scalar functions. In this case we write
\begin{equation}
r^{s}_k(x)=\frac{R^{s}_k(x)}{(1-x)^a x^b_{\nu_\alpha} \lambda^c(x,x_{\nu_\alpha})},
\end{equation}
where $\lambda(u,v)=1+u^2+v^2-2(u+v+uv)$, whereas the triads of integers $(a,\,b,\,c)$ are $(4,\,2,\,0)$ for $f^s_k(x)$, $(3,\,1,\,0)$ for $g^s_k(x)$, and $(1,\,2,\,2)$ for $\hat f^s_k(x)$ and $\hat g^s_k(x)$. Note that the $R^{s}_k(x)$ functions are not the same as those of Eq. \eqref{fsk}, but we use the same notation for simplicity. We also introduce the following dimensionless ultraviolet finite functions $\Delta_i(x)$
\begin{align}
\Delta_1(x)&=B_0(0, x\,m_{\rm LQ}^2, x\,m_{\rm LQ}^2)- B_0(0, \,m_{\rm LQ}^2,
m_{\rm LQ}^2)=-\log(x),\\
\Delta_2(x)&=B_0(x_{\nu_\alpha} m_{\rm LQ}^2, x\,m_{\rm LQ}^2, m_{\rm LQ}^2)- B_0(0, x\,m_{\rm LQ}^2, m_{\rm LQ}^2),
\end{align}
where the arguments of the two-point Passarino-Veltman scalar functions $B_0$ are given in the usual notation \cite{Mertig:1990an}. Note that the following two  three-point scalar functions appear throughout the calculation
\begin{align}
C_1(x)&=m_{\rm LQ}^2C_0(m_{\nu_\alpha}^2,m_{\nu_\alpha}^2,0,m_{\rm LQ}^2,x m_{\rm LQ}^2,m_{\rm LQ}^2),\\
C_2(x)&=m_{\rm LQ}^2C_0(m_{\nu_\alpha}^2,m_{\nu_\alpha}^2,0,x m_{\rm LQ}^2, m_{\rm LQ}^2,x m_{\rm LQ}^2),
\end{align}
but they can be written in terms of two-point scalar functions \cite{Devaraj:1997es} as follows
\begin{align}
C_1(x)&=\frac{1}{(x-1)  \lambda(x,x_{\nu_\alpha})}\Big((x-1) \left(x_{\nu_\alpha}+x-1\right)-x\left(x_{\nu_\alpha}-x+1\right)\Delta _1(x)-(x-1) \left(x_{\nu_\alpha}+x-1\right) \Delta _2(x)\Big),\\
C_2(x)&=\frac{1}{(x-1)\lambda(x,x_{\nu_\alpha})}\Big((x-1) \left(x_{\nu_\alpha}-x+1\right)- \left(x_{\nu_\alpha}+x-1\right)\Delta _1(x)-(x-1) \left(x_{\nu_\alpha}-x+1\right) \Delta _2(x)\Big).
\end{align}
It is also worth noticing that this calculation requires derivatives of the two- and three-point scalar functions, which we denote as
\begin{align}
C'_1(x)&=m_{\rm LQ}^4C'_0(m_{\nu_\alpha}^2,m_{\nu_\alpha}^2,q^2,m_{\rm LQ}^2,x m_{\rm LQ}^2,m_{\rm LQ}^2)|_{q^2=0},\\
C'_2(x)&=m_{\rm LQ}^4C'_0(m_{\nu_\alpha}^2,m_{\nu_\alpha}^2,q^2,x m_{\rm LQ}^2, m_{\rm LQ}^2,x m_{\rm LQ}^2)|_{q^2=0},
\end{align}
where the prime represents derivative with respect to $q^2$. Again  the $C'_i(x)$ functions can be decomposed in terms of two-point scalar functions as follows \cite{Devaraj:1997es}
\begin{align}
C'_1(x)&=
\frac{1}{6 (x-1) x_{\nu_\alpha}\lambda^2(x,x_{\nu_\alpha})}
\Big((x-1) x_{\nu_\alpha} \left(x^2 \left(3 x_{\nu_\alpha}-1\right)-3 x \left(x_{\nu_\alpha}^2-1\right)+\left(x_{\nu_\alpha}-1\right){}^3-x^3\right)\nonumber\\&+x  x_{\nu_\alpha} \left(4 x \left(x_{\nu_\alpha}+1\right)-\left(x_{\nu_\alpha}-1\right){}^2-3 x^2\right)\Delta _1(x)-(x-1)  \left(x_{\nu_\alpha}+x-1\right) \left(
\lambda(x,x_{\nu_\alpha})-2 x x_{\nu_\alpha}\right)\Delta _2(x)\Big),\\
C'_2(x)&=
\frac{1}{6 (x-1) x_{\nu_\alpha}\lambda^2(x,x_{\nu_\alpha})}
\Big((x-1) x_{\nu_\alpha} \left(3 x^2 \left(x_{\nu_\alpha}+1\right)-3 x x_{\nu_\alpha}^2-\left(x_{\nu_\alpha}-1\right){}^3-x^3-x\right)\nonumber\\&
+x  x_{\nu_\alpha} \left(2 x \left(x_{\nu_\alpha}+2\right)-x_{\nu_\alpha}^2+4 x_{\nu_\alpha}+x^2-3\right)\Delta _1(x)-(x-1) x  \left(x_{\nu_\alpha}-x+1\right) \left(
\lambda(x,x_{\nu_\alpha})-2  x_{\nu_\alpha}\right)\Delta _2(x)\Big).
\end{align}
We also need the following two-point scalar function derivative
\begin{equation}
B'_0(q^2,m_a^2,m_a^2)=\frac{1}{6m_a^2}.
\end{equation}

Below we present the $R^s_k(x)$ functions below for the contributions of scalar and vector LQs.

\subsection{Scalar LQ contribution}
For the contribution of scalar LQs to the static neutrino MDM and EDM we have the following $F^0_k(x)$ and $G^0_k(x)$ functions
\begin{align}
F^0_1(x)&=-\frac{1}{2} \Big(
-\left(x-1\right) x_{\nu_\alpha} \left(2 \left(3 x+1\right) x_{\nu_\alpha}+\left(x+1\right) \left(x-1\right){}^2\right)-2  x x_{\nu_\alpha} \left(\left(x+3\right) x_{\nu_\alpha}+\left(x-1\right){}^2\right)\Delta _1(x)\nonumber\\&+2  \left(x-1\right) \left(\left(x-1\right){}^2 x_{\nu_\alpha}+\left(3 x+1\right) x_{\nu_\alpha}^2+\left(x-1\right){}^4\right)\Delta _2(x)\Big),\\
F^0_2(x)&=-\frac{1}{2}\Big(\left(x-1\right) x_{\nu_\alpha} \left(2 x \left(x+3\right) x_{\nu_\alpha}+\left(x+1\right) \left(x-1\right){}^2\right)+2  x x_{\nu_\alpha} \left(x \left(x+3 x_{\nu_\alpha}-2\right)+x_{\nu_\alpha}+1\right)\Delta _1(x)\nonumber\\&- \left(2 x \left(x-1\right){}^3 x_{\nu_\alpha}+2 x \left(x+3\right) \left(x-1\right) x_{\nu_\alpha}^2+2 \left(x-1\right){}^5\right)\Delta _2(x)\Big),
\end{align}
and
\begin{align}
G^0_1(x)&=-2\left(-2 \left(x-1\right)  x_{\nu_\alpha}- \left(x+1\right) x_{\nu_\alpha}\Delta _1(x)+ \left(x-1\right) \left(\left(x-1\right){}^2+2 x_{\nu_\alpha}\right)\Delta _2(x)\right),\\
G^0_2(x)&=-2\left(\left(x^2-1\right) x_{\nu_\alpha}+2  x x_{\nu_\alpha}\Delta_1(x)- \left(x-1\right)  \left(x \left(x+x_{\nu_\alpha}-2\right)+x_{\nu_\alpha}+1\right)\Delta _2(x)\right).
\end{align}

For small $x_{\nu_\alpha}$ we obtain the following expansion to order $O\left(x_{\nu_\alpha}^2\right)$
\begin{align}
\Delta_2(x)&=\frac{1}{6 \left(x-1\right){}^5}\Big(
3 \left(x-1\right){}^2 \left(x^2-2 x \log \left(x\right)-1\right)x_{\nu_\alpha}+
 \left(x \left(x \left(x+9\right)-6(1+x)\log(x)-9\right)-1\right)x_{\nu_\alpha}^2\Big)\nonumber\\&+O\left(x_{\nu_\alpha} ^4\right),
\end{align}
which after their substitution  in the $r^0_k(x)$ functions yield the results presented in Eqs. \eqref{ffunc1}, \eqref{ffunc2}, \eqref{gfunc1}, and \eqref{gfunc2} at leading order in the neutrino mass.

As far as the scalar LQ contributions to the effective NCR are concerned, we obtain the following   $\hat F^0_k(x)$ and $\hat G^0_k(x)$ functions
\begin{align}
\hat F^0_1(x)&=\frac{1}{4} \Big((x-1) \left(x-1-x_{\nu_\alpha}\right) x_{\nu_\alpha} \left(x_{\nu_\alpha}^3-7 (x+1) x_{\nu_\alpha}^2+(x (11 x+26)+11) x_{\nu_\alpha}-5 (x-1)^2 (x+1)\right)\nonumber\\&
-2  x_{\nu_\alpha} \Big(5 x^4-15 x^3 \left(x_{\nu_\alpha}+1\right)+x^2 \left(x_{\nu_\alpha} \left(17 x_{\nu_\alpha}+4\right)+15\right)+x \left(x_{\nu_\alpha} \left(13-9 x_{\nu_\alpha}^2+x_{\nu_\alpha}\right)-5\right)\nonumber\\&+2 \left(x_{\nu_\alpha}-1\right){}^3 x_{\nu_\alpha}\Big)\Delta _1(x)+2 (x-1)  \Big(\left(1-26 x^2+x\right) x_{\nu_\alpha}^3-(20 x+13) (x-1)^3 x_{\nu_\alpha}\nonumber\\&+(x (32 x+5)+11) (x-1) x_{\nu_\alpha}^2-2 x_{\nu_\alpha}^5+(11 x+4) x_{\nu_\alpha}^4+5 (x-1)^5\Big)\Delta _2(x)\Big),\\
\hat F^0_2(x)&=\frac{1}{4}
\Big((x-1) \left(x-1-x_{\nu_\alpha}\right) x_{\nu_\alpha} \left(x_{\nu_\alpha}^3-7 (x+1) x_{\nu_\alpha}^2+(x (11 x+26)+11) x_{\nu_\alpha}-5 (x-1)^2 (x+1)\right)-2  x_{\nu_\alpha}\nonumber\\&
\times \Big(5 x^4-15 x^3 \left(x_{\nu_\alpha}+1\right)+x^2 \left(x_{\nu_\alpha} \left(17 x_{\nu_\alpha}+4\right)+15\right)+x \left(x_{\nu_\alpha} \left(13-9 x_{\nu_\alpha}^2+x_{\nu_\alpha}\right)-5\right)+2 \left(x_{\nu_\alpha}-1\right){}^3 x_{\nu_\alpha}\Big)\Delta _1(x)\nonumber\\&+
2 (x-1) \Big(\left(1-26 x^2+x\right) x_{\nu_\alpha}^3-(20 x+13) (x-1)^3 x_{\nu_\alpha}+(x (32 x+5)+11) (x-1) x_{\nu_\alpha}^2-2 x_{\nu_\alpha}^5\nonumber\\&+(11 x+4) x_{\nu_\alpha}^4+5 (x-1)^5\Big)\Delta _2(x)\Big),
\end{align}
and
\begin{align}
\hat G^0_1(x)&=
2 (x-1)  \left(x_{\nu_\alpha}-x-1\right) x_{\nu_\alpha} \left(x_{\nu_\alpha}^2-2 (x+2) x_{\nu_\alpha}+(x-1)^2\right)\nonumber
\\&+2   x_{\nu_\alpha} \left(x_{\nu_\alpha}^3-2 (2 x+1) x_{\nu_\alpha}^2+x (5 x+6) x_{\nu_\alpha}+x_{\nu_\alpha}-2 (x-1)^2 x\right)\Delta _1(x)\nonumber\\&
+2 (x-1)   \left(3 \left(3 x^2+1\right) x_{\nu_\alpha}^2-(7 x+5) (x-1)^2 x_{\nu_\alpha}+x_{\nu_\alpha}^4-(5 x+1) x_{\nu_\alpha}^3+2 (x-1)^4\right)\Delta _2(x),
\\
\hat G^0_2(x)&=
-2 (x-1)\left(-x_{\nu_\alpha}+x+1\right) x_{\nu_\alpha} \left(x_{\nu_\alpha}^2-2 (x+2) x_{\nu_\alpha}+(x-1)^2\right)\nonumber\\&+2  x_{\nu_\alpha} \left(x_{\nu_\alpha}^3-2 (2 x+1) x_{\nu_\alpha}^2+x (5 x+6) x_{\nu_\alpha}+x_{\nu_\alpha}-2 (x-1)^2 x\right)\Delta _1(x)\nonumber\\&+
2 (x-1)  \left(3 \left(3 x^2+1\right) x_{\nu_\alpha}^2-(7 x+5) (x-1)^2 x_{\nu_\alpha}+x_{\nu_\alpha}^4-(5 x+1) x_{\nu_\alpha}^3+2 (x-1)^4\right)\Delta _2(x).
\end{align}

\subsection{Vector LQ contribution}
As for the vector contribution to the neutrino MDM and EMD, the $F^1_k(x)$ and $G^1_k(x)$ functions are given by
\begin{align}
F^1_1(x)&=\frac{1}{4} \Big(
(x-1) x_{\nu_\alpha} \left((x (x (x+5)+21)-3) x_{\nu_\alpha}+(x+1) (x+2) (x-1)^2\right)
\nonumber\\&+2 x  x_{\nu_\alpha} \left(x \left(11 x_{\nu_\alpha}-3\right)+x_{\nu_\alpha}+x^3+2\right)\Delta _1(x)\nonumber\\&
+ \left(2 ((x-5) x+1) (x-1)^3 x_{\nu_\alpha}-2 (x (4 x+9)-1) (x-1) x_{\nu_\alpha}^2-2 (x+2) (x-1)^5\right)
\Delta _2(x)\Big),\\
F^1_2(x)&=\frac{1}{4} \Big(
-(x-1) x_{\nu_\alpha} \left(x \left(x \left((25-x) x_{\nu_\alpha}+x(1+x)-3\right)-x_{\nu_\alpha}-1\right)+x_{\nu_\alpha}+2\right)\nonumber\\&
-2 x x_{\nu_\alpha} \left(x \left((4 x+9) x_{\nu_\alpha}+x^2-3\right)-x_{\nu_\alpha}+2\right)\Delta _1(x)
\nonumber
\\&+ \left(6 (2 x-1) (x-1)^3 x_{\nu_\alpha}+2 x (11 x+1) (x-1) x_{\nu_\alpha}^2+2 (x+2) (x-1)^5\right)\Delta _2(x)\Big),
\end{align}
and
\begin{align}
G^1_1(x)&=
\left(3-x(1+x+x^2)\right)  x_{\nu_\alpha}
-(x+5) x x_{\nu_\alpha}\Delta _1(x)\nonumber\\&+(x-1) \left(x \left((6-x) x_{\nu_\alpha}+x^2-3\right)+x_{\nu_\alpha}+2\right)\Delta _2(x),\\
G^1_2(x)&=
(x-1) (7 x-1) x_{\nu_\alpha}+
2 (2 x+1) x  x_{\nu_\alpha}\Delta _1(x)-(x-1)   \left(x \left(7 x_{\nu_\alpha}-3\right)-x_{\nu_\alpha}+x^3+2\right)\Delta _2(x),
\end{align}
where the  $\Delta_i(x)$ functions were defined above, but with $m_{\rm LQ}=m_V$ instead.

The contribution of gauge LQs to the effective NCR is obtained via the $\hat F^1_k(x)$  functions only given our assumptions for the coupling constants and  can be written as
\begin{align}
\hat F^1_1(x)&=\frac{1}{2} \Big((x-1) x_{\nu_\alpha} \Big(2 x^3 \left(2 x_{\nu_\alpha}+5\right)+2 x^2 x_{\nu_\alpha} \left(9 x_{\nu_\alpha}+11\right)-2 x \left(x_{\nu_\alpha} \left(14 x_{\nu_\alpha}^2+x_{\nu_\alpha}+4\right)+5\right)
\nonumber\\&+\left(x_{\nu_\alpha}-1\right){}^2 \left(x_{\nu_\alpha} \left(11 x_{\nu_\alpha}-8\right)+5\right)-5 x^4\Big)-2  x_{\nu_\alpha} \left(x_{\nu_\alpha}+x-1\right) \Big(2 x_{\nu_\alpha}^3+(x-4) x_{\nu_\alpha}^2-2 x (4 x+9) x_{\nu_\alpha}\nonumber\\&+2 x_{\nu_\alpha}+5 (x-1)^2 x\Big)\Delta _1(x)+2 (x-1)  \Big((14 x+19) (1-x)^3 x_{\nu_\alpha}+(x (14 x+5)+29) (x-1) x_{\nu_\alpha}^2\nonumber\\&-2 x_{\nu_\alpha}^5+(5 x-2) x_{\nu_\alpha}^4+((13-8 x) x+19) x_{\nu_\alpha}^3+5 (x-1)^5\Big)\Delta _2(x)\Big),
\end{align}
\begin{align}
\hat F^1_2(x)&=\frac{1}{2}\Big(
(x-1) x_{\nu_\alpha} \left(10 x^3 \left(x_{\nu_\alpha}-1\right)-48 x^2 x_{\nu_\alpha} \left(x_{\nu_\alpha}+1\right)+2 x \left(x_{\nu_\alpha}-1\right){}^2 \left(19 x_{\nu_\alpha}+5\right)-5 \left(x_{\nu_\alpha}-1\right){}^4+5 x^4\right)\nonumber\\
&+2 x  x_{\nu_\alpha} \left(7 x_{\nu_\alpha}^2
-12 (x+1) x_{\nu_\alpha}+5 (x-1)^2\right) \left(x \left(x_{\nu_\alpha}+1\right)-\left(x_{\nu_\alpha}-1\right){}^2\right)\Delta _1(x)\nonumber\\&
-2 (x-1)  \Big(\left(-5 x^3+63 x-58\right) x_{\nu_\alpha}^2-3 (2 x+9) (x-1)^3 x_{\nu_\alpha}+7 x_{\nu_\alpha}^5-3 (4 x+11) x_{\nu_\alpha}^4+(x (11 x-1)+62) x_{\nu_\alpha}^3\nonumber\\&+5 (x-1)^5\Big)\Delta _2(x)\Big),
\end{align}
\begin{align}
\hat F^1_3(x)&=\frac{1}{4}\Big(2  x_{\nu_\alpha} \left(2 x_{\nu_\alpha}^3+(x-4) x_{\nu_\alpha}^2-2 x (4 x+9) x_{\nu_\alpha}+2 x_{\nu_\alpha}+5 (x-1)^2 x\right) \left(x \left(2 x_{\nu_\alpha}+1\right)-\left(x_{\nu_\alpha}-1\right) x_{\nu_\alpha}-x^2\right)\Delta _1(x)\nonumber\\&+2 (x-1) \big(-\left(x \left(30 x^2+26 x+5\right)+11\right) x_{\nu_\alpha}^3-2 x_{\nu_\alpha}^6+(5 x+4) x_{\nu_\alpha}^5+(x (5 x+9)+1) x_{\nu_\alpha}^4\nonumber\\&+(x-1) (x (8 x (5 x+3)-3)-13) x_{\nu_\alpha}^2-(x-1)^3 (x (23 x+15)-5) x_{\nu_\alpha}+5 (x-1)^5 x\big)\Delta _2(x)\nonumber\\&-(x-1) x_{\nu_\alpha} \big(x_{\nu_\alpha}^5+(x-6) x_{\nu_\alpha}^4-2 (x (7 x+10)-2) x_{\nu_\alpha}^3+2 (x (x (13 x+24)+32)+3) x_{\nu_\alpha}^2\nonumber\\&-(x-1) (x (x (19 x+31)+3)-5) x_{\nu_\alpha}+5 (x-1)^3 x (x+1)\big)\Big),
\end{align}
and
\begin{align}
\hat F^1_4(x)&=\frac{1}{4 } \Big(
(x-1) x_{\nu_\alpha} \big(-x^4 \left(17 x_{\nu_\alpha}+10\right)+2 x^3 x_{\nu_\alpha} \left(13 x_{\nu_\alpha}-1\right)+2 x^2 \left(-13 x_{\nu_\alpha}^3+7 x_{\nu_\alpha}^2+x_{\nu_\alpha}+5\right)\nonumber\\&+x \left(x_{\nu_\alpha}-1\right){}^2 \left(x_{\nu_\alpha}+1\right) \left(17 x_{\nu_\alpha}-5\right)-5 \left(x_{\nu_\alpha}-1\right){}^4 x_{\nu_\alpha}+5 x^5\big)\nonumber\\&
-2 x x_{\nu_\alpha} \big(x_{\nu_\alpha}^5-4 (x+2) x_{\nu_\alpha}^4+3 (x (2 x+5)+6) x_{\nu_\alpha}^3-(x (x (4 x+11)-7)+16) x_{\nu_\alpha}^2\nonumber\\&+(x (x (x+1) (x+8)-23)+5) x_{\nu_\alpha}-5 (x-1)^3 x\big)\Delta _1(x)\nonumber\\&
2 (x-1) \big(\left(x \left(6 x^2+x-17\right)+34\right) x_{\nu_\alpha}^3-3 \left(x \left(x \left(5 x^2+x+6\right)-19\right)+7\right) x_{\nu_\alpha}^2\nonumber\\&-x_{\nu_\alpha}^6+3 (x+3) x_{\nu_\alpha}^5-(x (3 x+10)+26) x_{\nu_\alpha}^4+(x-1)^3 (x (15 x+23)-5) x_{\nu_\alpha}-5 (x-1)^5 x\big)\Delta _2(x)\Big).
\end{align}

\section{LQ contributions to low-energy observables}
\label{lowenerobser}
In this Appendix we present the most relevant analytical expressions for the LQ contributions to the low-energy observables useful to constrain the LQ Yukawa couplings $\zeta^s_{L\,i\alpha}$ and $\zeta^s_{R\,i\alpha}$ of Eqs. \eqref{genSLag} and \eqref{genVLag} for the  LQs arisen from the scalar $\widetilde{R}_2$ and vector $U_1$ representations. We remind the reader that the index $i$ ($i=1,2$) corresponds to the quark family, whereas the index $\alpha$ ($\alpha=\mu,\,\tau$) corresponds to the lepton family. We  have also assumed that the couplings to the fermions of the first generation vanish. Such expressions can be found in several sources but the results shown below are those compiled in \cite{Becirevic:2016yqi,Becirevic:2016oho} for the scalar LQ contribution and \cite{Cornella:2019hct} for the vector LQ contribution. 
\subsection{$\widetilde{R}_2$ representation}
The $\widetilde{R}_2$  contribution to the Wilson coefficients relevant for the $b\to s\ell_1\ell_2$ decay in terms of the coupling constants of Eq. \eqref{genSLag} is
\begin{equation}
\left(C_9^{\ell_1\ell_2}\right)'=-\left(C_{10}^{\ell_1\ell_2}\right)'=-\frac{\pi\upsilon^2}{2V_{tb}V^*_{ts}\alpha_{\rm em}}\frac{\zeta^0_{L\,3\ell_1}\zeta^0_{L\,3\ell_2}}{m_{\rm LQ}^2}.
\end{equation}
For the mass difference of the $B_s-\overline{B}_s$ system we have
\begin{align}
\frac{\Delta m_{B_s}^{\rm Theor.}}{\Delta m_{B_s}^{\rm SM}}&=1+\frac{\eta_1}{16G_F^2m_W^2|V_{tb}V^*_{ts}|^2\eta_B S_0(x_t)m_{\rm LQ}^2}\Bigg(\left(\zeta^0_{L\,3\alpha}\zeta^{0*}_{L\,2\alpha}\right)^2+\frac{1}{2}\left(\zeta^0_{R\,3\alpha}\zeta^{0*}_{R\,2\alpha}\right)^2\nonumber\\&
-\eta_{41}\frac{3}{2}\left(\zeta^0_{L\,3\alpha}\zeta^{0*}_{L\,2\alpha}\right)\left(\zeta^0_{R\,3\alpha'}\zeta^{0*}_{R\,2\alpha'}\right)\left(\frac{m_{B_s}}{m_b+m_s}\right)^2\frac{B_4(m_b)}{B_1(m_b)}\Bigg),
\end{align}
where again repeated indices $\alpha$ and $\alpha'$ sum over $\mu$ and $\tau$,  $x_t=m_t^2/m_W^2$, and the $S_0(x)$ is Inami-Lim function
\begin{equation}
S_0(x)=\frac{1}{4}\left(1+\frac{9}{1-x}-\frac{6}{(1-x)^2}-\frac{6x^2}{(1-x)^3}\log(x)\right).
\end{equation}
Also, we  use $\eta_1=0.81(1)$, $\eta_{41}=4.4(1)$ \cite{Becirevic:2016yqi}, whereas  $B_1(m_b)=3.4$ and $B_4(m_b)=4.5$ for the bag parameters \cite{Aoki:2016frl}. 

The corresponding  contribution to the $B\to K\nu\nu$ decay can be written as
\begin{equation}
R_{\nu\nu}=\frac{{\rm BR}(B\to K\nu\nu)^{\rm Theor.}}{{\rm BR}(B\to K\nu\nu)^{\rm SM}}=1-\frac{1}{6C_L^{\rm SM}}{\rm Re}\left(\frac{\zeta^0_{L\,3\alpha}\zeta^{0*}_{L\,3\alpha}}{Nm_{\rm LQ}^2}\right)+\frac{1}{48{C_L^{\rm SM}}^2}\frac{\zeta^0_{L\,3\alpha}\zeta^{0*}_{L\,3\alpha}\zeta^0_{L\,3\alpha'}\zeta^{0*}_{L\,3\alpha'}}{|N|^2m_{\rm LQ}^4},
\end{equation}
where again repeated indices $\alpha$ and $\alpha'$ sum over $\mu$ and $\tau$, whereas
$N = \dfrac{G_F V_{tb} V_{ts}^*\alpha_{\rm em}}{\sqrt{2}\pi}$ and $C_L^{\rm SM}=-6.38(10)$ \cite{Brod:2010hi}.

The scalar LQ contributions to the LFV decays $\tau\to \mu\gamma$ and $\tau\to \phi\gamma$ can be written as \cite{Becirevic:2016oho}
\begin{equation}
{\rm BR}(\tau\to \mu\gamma)=\frac{\alpha_{\rm em}(m_\tau^2-m_\mu^2)^3}{4m_\tau^3\Gamma_\tau}\left|\zeta^0_{L\,3\mu}\zeta^{0*}_{L\,3\tau}\frac{N_c m_b^2  m_\tau}{96\pi^2 m_{\rm LQ}^4}\left(\frac{5}{2}+\log\left(\frac{m_b^2}{m_{\rm LQ}^2}\right)\right)\right|^2,
\end{equation}
and
\begin{equation}
{\rm BR}(\tau\to \mu\phi)=\frac{f_\phi^2m_\phi^4}{256\pi m_\tau^3\Gamma_\tau}\left|\frac{\zeta^0_{L\,2\tau}\zeta^{0*}_{L\,2\mu}}{m_{\rm LQ}^2}\right|^2\left(-1+\frac{m_\mu^2+m_\tau^2}{2m_\phi^2}+\frac{\left(m_\mu^2-m_\tau^2\right)^2}{2m_\phi^4}\right)\lambda^{1/2}(m_\phi^2,m_\tau^2,m_\mu^2),
\end{equation}
where the triangle function is $\lambda(x,y,z)=x^2+y^2+z^2-2(xy+xz+yz)$ and the $\phi$ decay constant $f_\phi=241(8)$ MeV.

\subsection{$U_1$ representation}
We consider the model of Ref. \cite{DiLuzio:2017vat} where the LQ coupling constants of Eq. \eqref{genVLag} are given by $\zeta^1_{L\,\i\alpha}=g_4/\sqrt{2}\beta_{L\,i\alpha}$ and  $\zeta^1_{R\,\i\alpha}=0$, with $g_4=2 m_{\rm LQ}\sqrt{C_U}/\upsilon$.  Below we list the $U_1$ contribution to some observables that can be useful to constrain the LQ couplings and fit the model parameters.

The $U_1$ contribution to the $B\to D\tau\nu$ and $B\to D^*\tau\nu$  decays reads as
\begin{equation}
R_D=\frac{{\rm BR}\left(B\to D\tau\nu\right)}{{\rm BR}\left(B\to D\tau\nu\right)_{\rm SM}}\simeq R_{D^*}\simeq R_D^{\rm SM}\left(1+2 C_U {\rm Re}\left(1+\frac{V_{cs}}{V_{cb}}\beta_{L\,2\tau}\right)\right).
\end{equation}
For the decay $b\to s\ell^-\ell^+$ we have
\begin{equation}
C_9^{\ell\ell}=-C_{10}^{\ell_\ell}=-\frac{2\pi}{V_{tb}V^*_{ts}\alpha_{\rm em}}C_U \beta_{L\,2\ell}\beta^*_{L\,3\ell}.
\end{equation}
The contribution to the following $B$ meson decays can also be useful to constrain the $U_1$ LQ coupling constants
\begin{equation}
{\rm BR}\left(B\to K\tau^- \tau^+\right)\simeq 1.5\times 10^{-7}+1.4\times 10^{-3} C_U{\rm Re}\left(\beta_{L\,2\tau}\right)+3.5 C_U^2\left|\beta_{L\,2\tau}\right|^2,
\end{equation}
\begin{equation}
{\rm BR}\left(B^+\to K^+\tau^+ \mu^-\right)=8.3C_U^2 \left|\beta_{L\,2\mu}\right|^2,
\end{equation}
\begin{equation}
{\rm BR}\left(B^+\to K^+\tau^- \mu^+\right)=8.3C_U^2 \left|\beta_{L\,3\mu}\beta_{L\,2\tau}^*\right|^2,
\end{equation}
and
The $U_1$ contribution to the LFV tau decay $\tau\to \mu\phi$ is given by
\begin{equation}
{\rm BR}(\tau\to \mu\phi)=\frac{f_\phi^2 G_F^2}{16\pi \Gamma_\tau}m_\tau^3\left(1-\frac{m_\phi^2}{m_\tau^2}\right)^2
\left(1+2\frac{m_\phi^2}{m_\tau^2}\right)C_U^2\left|\beta_{L\,2\tau}\beta_{L\,2\mu}^*\right|^2.
\end{equation}
The masses and life times of the $B$ mesons as well as the latest experimental constraints and measurements on the $B$-meson decay modes were taken from \cite{ParticleDataGroup:2022pth} and \cite{HeavyFlavorAveragingGroup:2022wzx}.

\end{document}